\documentclass[aps,prb,twocolumn,floatfix]{revtex4-1}

\usepackage{amsmath}
\usepackage{amsfonts}
\usepackage{graphicx}
\usepackage{float}
\usepackage{dcolumn}
\usepackage{bm}
\usepackage{xcolor}
\usepackage{subfigure}
\usepackage{comment}

\begin{document}

\title{Tunable zero-energy Dirac and Luttinger nodes in a two-dimensional topological superconductor}

\author{Ryan Mays}
\affiliation{Department of Physics and Astronomy, George Mason University, Fairfax, VA 22030, USA}

\author{Predrag Nikoli\'c$^{1,2}$}
\affiliation{$^1$Department of Physics and Astronomy, George Mason University, Fairfax, VA 22030, USA}
\affiliation{$^2$Institute for Quantum Matter at Johns Hopkins University, Baltimore, MD 21218, USA}

\date{\today}

\begin{abstract}

Cooper pairing in ultrathin films of topological insulators, induced intrinsically or by proximity effect, can produce an energetically favorable spin-triplet superconducting state. The spin-orbit coupling acts as an SU(2) gauge field and stimulates the formation of a spin-current vortex lattice in this superconducting state. Here we study the Bogoliubov quasiparticles in such a state and find that the quasiparticle spectrum consists of a number of Dirac nodes pinned to zero energy by the particle-hole symmetry. Some nodes are ``accidental'' and move through the first Brillouin zone along high-symmetry directions as the order parameter magnitude or the strength of the spin-orbit coupling are varied. At special parameter values, nodes forming neutral quadruplets merge and become gapped out, temporarily producing a quadratic band-touching spectrum. All these features are tunable by controlling the order parameter magnitude via a gate voltage in a heterostructure device. In addition to analyzing the spectrum at the mean-field level, we briefly discuss a few experimental signatures of this spectrum.

\end{abstract}

\maketitle

\section{Introduction}

Topological materials can exhibit many unconventional physical phenomena, from symmetry-protected gapless states at system boundaries, to exotic correlation effects resulting in topological order \cite{WenQFT2004}. Even though such phenomena are normally associated with a gap of the bulk spectrum, topologically non-trivial dynamics can also occur in the systems without a bulk gap. The best known example are Weyl semimetals \cite{Ari2010, Burkov2011a}, experimentally identified in several materials \cite{Armitage2018}. A Weyl spectrum features three-dimensional massless chiral quasiparticles in which the intrinsic spin degeneracy is not lifted only at special nodal-point wavevectors in the first Brillouin zone. Then, the bulk topological properties are most strikingly revealed when the chemical potential sits exactly at the energy of the Weyl nodes, or in close proximity -- so that the low-energy excitations are strongly affected by the nodal singularities of the band's Berry curvature. In reality, finding materials that satisfy this condition has proven hard, and tuning the chemical potential in three-dimensional systems can be achieved only by doping, which unfortunately introduces disorder or alters the crystal properties.

Two-dimensional systems are much more tunable in this regard. The chemical potential of a two-dimensional electron gas embedded in a heterostructure device can be routinely controlled by a gate voltage. However, it is more difficult to obtain massless chiral quasiparticles in this setting. Surfaces of strong topological insulators do naturally host such quasiparticles \cite{Fu2007, Moore2007, Teo2008}, but their existence still relies on the presence of a three-dimensional insulating bulk which spoils the gate control. Ultrathin films of topological insulators can be made, but their surface states become gapped due to geometric proximity. Graphene hosts massless degenerate (non-chiral) Dirac electrons, which would become gapped in favor of a topological insulator if the spin-orbit coupling were sizeable \cite{Kane2005}.

\begin{figure}[t]
\centering
        \includegraphics[width=0.3\textwidth]{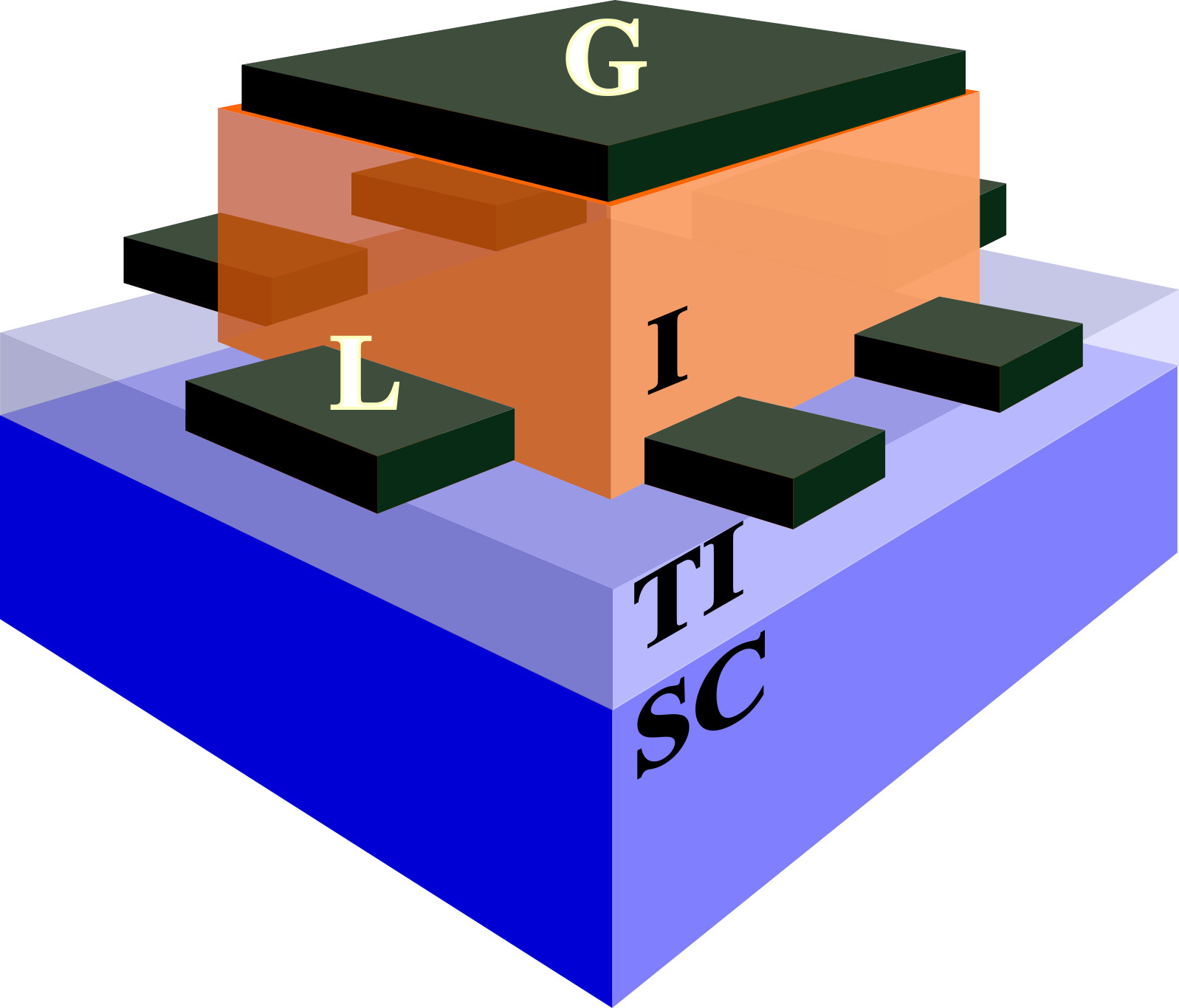}
        \caption{\label{Heterostructure}A gated heterostructure device which induces Cooper pairing inside an ultra-thin topological insulator (TI) film, proposed in \cite{Nikolic2011a}. The TI film is placed in contact with a conventional superconductor (SC) and separated from a top gate (G) by an insulating layer (I). Manipulating the gate voltage can then drive superconductor-insulator transitions inside the TI, and the properties of the correlated electron states inside the TI can be explored by transport measurements through the attached leads (L)}
\end{figure}

Here we explore a different approach to creating massless nodal quasiparticles in tunable two-dimensional systems. This approach may be hard to realize in practice, but it is theoretically sound. It relies on the ability to generate via proximity effect a superconducting state in an ultrathin film made from a topological insulator material. This is theoretically possible: a conventional phonon superconductor with a sufficiently high critical temperature can overcome a small bandgap of the TI film and introduce Cooper pairing in the film via a non-mean-field proximity effect \cite{Nikolic2012b}. Then, the natural strong spin-orbit coupling of the Rashba-type inside the film gives spin-triplet Cooper pairs an advantage at large wavevectors, stimulating a non-uniform condensate \cite{Nikolic2011a, Nikolic2012a}. A heterostructure device which could realize these conditions is shown in Figure \ref{Heterostructure}. Mean-field calculations have indeed identified a prominent stable spin-triplet condensate which hosts a vortex lattice of spin currents and respects the time-reversal symmetry \cite{Nikolic2014a}. A physical intuition for such a state can be gained from a comparison with $s$-wave superconductors in magnetic fields. An external gauge field is known to promote the appearance of Abrikosov vortex lattices in type-II superconductors. By analogy, the spin-orbit coupling within a TI ultrathin film is mathematically equivalent to an SU(2) gauge field coupled to spin instead of charge currents \cite{Frohlich1992}. This SU(2) flux cannot be expelled (the system is of type-II), so a spin-triplet superconducting state is likely going to respond to it by developing a vortex lattice.

Another system in which a spin-current vortex lattice could potentially arise is the surface of a topological Kondo insulator. This requires multiple Fermi pockets of helical Dirac quasiparticles on the system boundary \cite{Kang2013, Lu2013b, Nikolic2014b}. Certain channels of inter-pocket electron scattering, here facilitated by Coulomb interactions, have the same characteristics as the inter-surface coupling due to Cooper pairing in the previous ultrathin film system. If a vortex lattice were stabilized, it would exist within a neutral exciton condensate so the quasiparticle charge dynamics would be much better exposed. No states of this kind have been found on the boundaries of topological Kondo insulators so far \cite{Nakajima2016, Takeuchi2016}, but perhaps it takes some material and interface engineering to create them.

In this paper, we analyze the Bogoliubov quasiparticles of the mentioned spin-triplet topological superconducting state in a TI ultrathin film. We show that the quasiparticle spectrum contains a number of massless Dirac points placed at different wavevectors in momentum space. These nodes are protected by the intrinsic and spontaneously broken symmetries despite band degeneracy. Furthermore, the particle-hole symmetry of the superconducting state ensures that the chemical potential is always exactly at the node energy. By gating this system, it is possible to control the magnitude of the order parameter inside the film. This in turn tunes the locations of some nodes in momentum space. There are special parameter values for which two pairs of opposite-chirality Dirac nodes meet at the same wavevector and temporarily form a quadratic band-touching spectrum before gapping out. This is a particularly interesting feature of the system, because the chemical potential is still at the ensuing quadratic Luttinger node, while the bands with quadratic band touching have an elevated density of states at the chemical potential and become susceptible to instabilities caused by Coulomb and other interactions. Therefore, this system may be a tunable testbed for novel strongly correlated states of matter with non-trivial topology. 

The prospect of three-dimensional quadratic band touching has recently attracted considerable attention in the context of in pyrochlore iridates \cite{Pesin2010, Chen2012, Ishii2015, Kondo2015, Nakayama2016, Zhang2017, Wang2017, Cheng2017, Shinaoka2019}. This kind of a nodal semimetallic state turns at least into a non-Fermi liquid phase due to Coulomb interactions \cite{Abrikosov1971, Moon2013}, or perhaps succumbs to further instabilities leading to localization in the presence of disorder \cite{Parameswaran2017} or unconventional states with spontaneously broken symmetries \cite{Herbut2014, Herbut2015, Herbut2016b, Herbut2017, Herbut2017b, Herbut2016, Nevidomskyy2019}. The Fermi energy is not required to sit at the quadratic band-touching node in pyrochlore iridates. The hypothetical system we consider has at least the advantage of tunability and pinning the Fermi energy exactly to the node.

Our work is related to topological superconductivity as well \cite{Sato2017}. The system we explore is an example of a two-dimensional nodal superconductor in the DIII symmetry class \cite{Schnyder2008, Schnyder2013, Schnyder2014, Shiozaki2014}, but it differs from the commonly studied cases \cite{Norman1995, Sato2006, Roy2008, Qi2009b, Norman2009, Fujimoto2009, Tanaka2009b, Kobayashi2014, Zhang2014, Zhao2016, Norman2017, Ikeda2017} by having a multi-component order parameter which breaks translational symmetry and antisymmetrizes its spin-triplet Cooper pair wavefunctions via an internal two-state degree of freedom (film surface index) instead of coordinates (like a $p$-wave superconductor). A two-fold degeneracy of the Bogoliubov quasiparticle bands and nodes is present in our model and protected by the time-reversal symmetry. Further topological protection is established only locally in momentum space by the association of topological indices to the nodes. At least superficially, the nodes carry a Z topological index \cite{Beri2010} which represents the winding number of the local quasiparticles' spin-momentum-locking texture. Perturbations which respect the time-reversal symmetry may reduce this to a Z$_2$ index \cite{Schnyder2014}, but we do not analyze it in the present work. Additional symmetries of the superconducting state in our system protect and pin a set of nodes to the high-symmetry wavevectors of the first Brillouin zone, while other nodes are ``accidental'' and migrate through momentum space as the model parameters are varied. It should be noted that the quasiparticle spectrum we discuss is distinct from Majorana modes found in some other topological superconductors.

The paper is organized as follows. We begin by presenting the model of our system in Section \ref{secModel}. Section \ref{secResults} presents all findings related to the topological insulator normal state (\ref{secNormal}), node protection mechanisms (\ref{secProtection}), the migration of ``accidental'' nodes (\ref{secMigration}), and the node merger into quadratic band touching points (\ref{secQuadratic}). Section \ref{secExperiments} outlines the anticipated experimental manifestations of the state we consider, and Section \ref{secConclusions} summarizes all conclusions.

\section{Model}\label{secModel}

As a model system, we consider electrons in an ultra-thin film made from a topological insulator (TI) material. The TI's surface states are gapped by the mutual proximity of the opposite surfaces, but remain the lowest energy states in the electron spectrum. A tight-binding Hamiltonian representing this system is:
\begin{equation}\label{H0}
    H_0 =
    \sum_i \left[-t\sum_{j \in i} \psi^\dagger_i (e^{-i \tau^z A_{ij}} + \Delta_{ij}\tau^x) \psi^{\phantom{\dagger}}_j -\mu \psi^\dagger_i\psi^{\phantom{\dagger}}_i \right] \ .
\end{equation}
The spinor operators $\psi_i^{\phantom{\dagger}}$ ($\psi_i^\dagger$) are fermionic annihilation (creation) operators in second quantization which generate electron hopping between the sites $i$ and their nearest neighbors $j \in i$ on the square lattice. The spin orbit coupling of strength $\alpha$ is encoded with a SU(2) gauge field $A_{ij} = -A_{ji}$ defined on the lattice links
\begin{eqnarray}\label{Gauge}
 A_{i,i+\hat{x}} &=& \alpha \sigma^y \\
 A_{i,i+\hat{y}} &=& -\alpha \sigma^x \ . \nonumber
\end{eqnarray}
This gauge field contains $\sigma^a$ ($a\in\lbrace x,y,z \rbrace$) Pauli matrices that couple to spin, and by itself produces massless Dirac excitations at every high-symmetry wavevector of the first Brillouin zone (1BZ). The $\tau^a$ Pauli matrices operate on the surface index represented by the $\pm$ (top/bottom) eigenvalue of $\tau^z$. Then, the $\Delta_{ij}\tau^x$ term describes intersurface tunneling. Three tunneling parameters are needed to control the independent gaps at the $\Gamma$, X and M points of the square lattice 1BZ:
\begin{eqnarray}\label{Deltas}
&& \sum_{\delta \bf{r}} \Delta_{\delta \bf{r}} e^{i \bf{k} \delta \bf{r}} 
  = \Delta_{\Gamma} + \Delta_{\textrm{M}} \sin^2\left(\frac{k_x}{2} \right) \sin^2\left(\frac{k_y}{2} \right) \\
&& ~ +\Delta_{\textrm{X}} \left[\sin^2\left(\frac{k_x}{2}\right) \cos^2\left(\frac{k_y}{2}\right) + \cos^2\left(\frac{k_x}{2} \right) \sin^2\left(\frac{k_y}{2}\right) \right] \nonumber \ ,
\end{eqnarray}
where $\delta{\bf r} = {\bf r}_i - {\bf r}_j$ in $\Delta_{\delta{\bf r}} \equiv \Delta_{ij}$.

\begin{figure}[t]
        \includegraphics[width=0.30\textwidth]{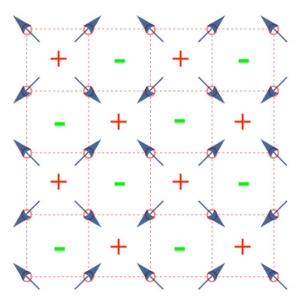}
        \caption{\label{EtaPattern}The spin-triplet superconducting order parameter $\bar{\eta} (e^{i\theta}, 0,  e^{-i\theta})$ featuring a spin-current vortex lattice. The arrows centered at microscopic lattice sites represent the spatial variations of the phase $\theta$. Since the gradients of $\theta$ determine spin currents, this amounts to a checkerboard pattern of vortices (+) and antivortices (-) on lattice plaquettes. The unit-cell of the superconducting state contains 2x2 plaquettes of the microscopic lattice.}
\end{figure}

To introduce superconductivity, we add to $H_0$ the following mean-field pairing Hamiltonian
\begin{align}\label{HI}
  H_{\textrm{int}}&=
     \sum_i \Bigl[U_t(|\eta_{i\resizebox{0.15cm}{0.15cm}{{$\uparrow$}}}|^2 + |\eta_{i\resizebox{0.15cm}{0.15cm}{{$\downarrow$}}}|^2) + U_{t0}|\eta_{i0}|^2\\
     &+\eta^{*}_{i\resizebox{0.15cm}{0.15cm}{{$\uparrow$}}} \psi_{i\resizebox{0.15cm}{0.15cm}{{$\uparrow$}}+}^{\phantom{*}}
     \psi_{i\resizebox{0.15cm}{0.15cm}{{$\uparrow$}}-}^{\phantom{*}} +
     \eta^*_{i\resizebox{0.15cm}{0.15cm}{{$\downarrow$}}} \psi_{i\resizebox{0.15cm}{0.15cm}{{$\downarrow$}}+}^{\phantom{*}}
     \psi_{i\resizebox{0.15cm}{0.15cm}{{$\downarrow$}}-}^{\phantom{*}} \nonumber\\ &+ \eta_{i0}^{\phantom{*}*}\frac{\psi_{i\resizebox{0.15cm}{0.15cm}{{$\uparrow$}}+}\psi_{i\resizebox{0.15cm}{0.15cm}{{$\downarrow$}}-} +\psi_{i\resizebox{0.15cm}{0.15cm}{{$\downarrow$}}+}\psi_{i\resizebox{0.15cm}{0.15cm}{{$\uparrow$}}-}}{\sqrt{2}}  + h.c. \Bigr] \nonumber \ ,
 \end{align}
 
\begin{figure}[]{
        \includegraphics[width=0.48\textwidth]{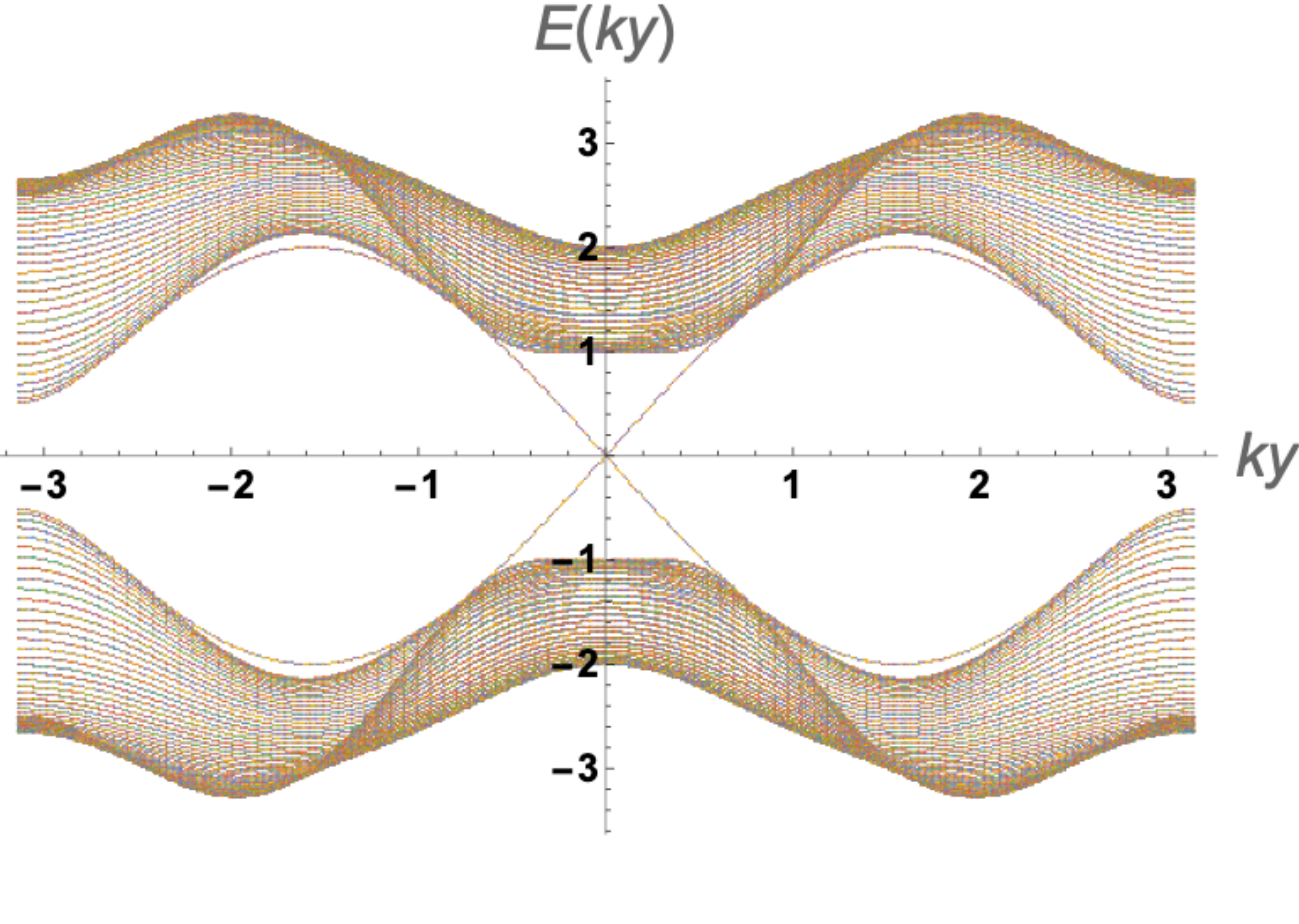}
        \caption{The energy spectrum achieved from diagonalizing \eqref{H0} with $\Delta_{\textrm{X}} = -2$ and $\Delta_{\Gamma} = 1$. A pair of edge states can be seen crossing the $\Gamma$ point.}\label{TISpectra}}
\end{figure}
\noindent{}which is a part of the full Hubbard-Stratonovich decoupling of all possible attractive short-range interactions among electrons \cite{Nikolic2011a} (presumably induced by phonons or a proximity effect \cite{Nikolic2012b}). We will consider only the spin triplet pairing channels, assuming that the spin-orbit coupling is strong. This is motivated by the fact that the Rashba-type spin-orbit coupling (\ref{Gauge}) is effectively a momentum-dependent Zeeman field, so bosonic Cooper pairs with a non-zero spin projection and large momentum can become the lowest energy collective excitations \cite{Nikolic2011a}. They can be softer even than the spin singlet pairs, which inherit the zero-momentum gap from the parent band-insulating electronic spectrum. If the spin triplet Cooper pairs condense at large wavevectors, they can form a time-reversal-invariant superconducting state with the order parameter
\begin{equation}\label{SCParam}
  {\eta}=
        \left( {\begin{array}{c}
        \eta_{\resizebox{0.15cm}{0.15cm}{{$\uparrow$}}}\\
        \eta_0\\
        \eta_{\resizebox{0.15cm}{0.15cm}{{$\downarrow$}}}\\
        \end{array} } \right)
  =
        \left( {\begin{array}{c} 
        \bar{\eta} e^{i\theta}\\
        0\\
        \bar{\eta} e^{-i\theta}
        \end{array}} \right) \ .
\end{equation}
The SU(2) flux of the gauge field (\ref{Gauge}) can impart a vortex lattice of spin currents in a spin triplet superconductor, in a manner similar to how an external magnetic field generates an Abrikosov lattice in conventional type-II superconductors. The topological defects of spin currents shaped by the Rashba spin-orbit coupling are peculiar and complex even in the continuum limit \cite{Nikolic2014}, requiring the presence of both vortices and antivortices of the winding phase $\theta$. On the square lattice, a dense checkerboard pattern of vortices and antivortices shown in Figure \ref{EtaPattern} was discovered in numerical mean-field calculations \cite{Nikolic2014a} as a stable superconducting phase in a broad parameter region with strong spin-orbit coupling. This particular state will be the starting point for our analysis.

We determine the quasiparticle spectrum in the triplet vortex lattice superconductor using the Bogoliubov-de Gennes (BdG) Hamiltonian constructed from (\ref{H0}) and (\ref{HI}). The presence of spin-triplet superconducting amplitudes requires an 8-component Nambu spinor per lattice site
\begin{equation}\label{Nambu}
  {\Psi^T_i} =
    (\begin{array}{cccccccc}
    \psi^{\phantom{*}}_{i+\resizebox{0.15cm}{0.15cm}{$\uparrow$}}
    &\psi^*_{i+\resizebox{0.15cm}{0.15cm}{$\downarrow$}}
    &\psi^{\phantom{*}}_{i+\resizebox{0.15cm}{0.15cm}{$\downarrow$}}
    &\psi^*_{i+\resizebox{0.15cm}{0.15cm}{$\uparrow$}}
    &\psi^{\phantom{*}}_{i-\resizebox{0.15cm}{0.15cm}{$\uparrow$}}
    &\psi^*_{i-\resizebox{0.15cm}{0.15cm}{$\downarrow$}}
    &\psi^{\phantom{*}}_{i-\resizebox{0.15cm}{0.15cm}{$\downarrow$}}
    &\psi^*_{i-\resizebox{0.15cm}{0.15cm}{$\uparrow$}} 
    \end{array})
\end{equation}
resulting with an unphysical doubling of the BdG spectrum that needs to be undone by hand. Without superconductivity that breaks the lattice translation symmetry, the BdG Hamiltonian in momentum space is:
\begin{widetext}
\begin{equation}\label{8x8}
H_0^{\textrm{(BdG)}} = \left( {\begin{array}{cccccccc}
    -2t C_\alpha C_{2k} & 0 & 2itS_\alpha S_{-2k} & 0 & \Delta(k_x,k_y) & 0 & 0 &  -\eta_{\resizebox{0.15cm}{0.15cm}{{$\uparrow$}}}\\
    0 & 2t C_\alpha C_{2k} & 0 & -2itS_\alpha S_{-2k} & 0 & -\Delta(k_x,k_y) & \eta^*_{\resizebox{0.15cm}{0.15cm}{{$\downarrow$}}} & 0\\
    -2itS_\alpha S_{+2k} & 0 & -2t C_\alpha C_{2k} & 0 & 0 & -\eta_{\resizebox{0.15cm}{0.15cm}{{$\downarrow$}}} & \Delta(k_x,k_y) & 0\\
    0 & 2itS_\alpha S_{+2k} & 0 & 2t C_\alpha C_{2k} & \eta^*_{\resizebox{0.15cm}{0.15cm}{{$\uparrow$}}} & 0 & 0 & -\Delta(k_x,k_y)\\
    \Delta(k_x,k_y) & 0 & 0 & \eta_{\resizebox{0.15cm}{0.15cm}{{$\uparrow$}}} & -2tC_\alpha C_{2k} & 0 & -2itS_\alpha S_{-2k} & 0 \\
    0 & -\Delta(k_x,k_y) & -\eta^*_{\resizebox{0.15cm}{0.15cm}{{$\downarrow$}}} & 0 & 0 & 2t C_\alpha C_{2k} & 0 & 2it S_\alpha S_{-2k}\\
    0 & \eta_{\resizebox{0.15cm}{0.15cm}{{$\downarrow$}}} & \Delta(k_x,k_y) & 0 & 2it S_\alpha S_{+2k} & 0 & -2t C_\alpha C_{2k} & 0\\
    -\eta^*_{\resizebox{0.15cm}{0.15cm}{{$\uparrow$}}} & 0 & 0 & -\Delta(k_x,k_y) & 0 & -2itS_\alpha S_{+2k} & 0 & 2t C_\alpha C_{2k}
   \end{array}} \right) 
\end{equation}
\end{widetext}
with the following terms defined as
\begin{eqnarray}
&C_\alpha = \cos(\alpha),\\  &S_\alpha = \sin(\alpha),\nonumber\\ &C_{2k} = \cos(2k_x) + \cos(2k_y),\nonumber\\  &S_{\pm2k} = \sin(2k_x) \pm i\sin(2k_y),\nonumber
\end{eqnarray}
and the intersurface tunneling is represented as
\begin{align}
\Delta(k_x,k_y)=& \,\Delta_{\Gamma} + \Delta_{\textrm{M}}\sin^2(k_x)\sin^2(k_y) \\
& +\Delta_{\textrm{X}} \left[\cos^2(k_x)\sin^2(k_y) + \sin^2(k_x)\cos^2(k_y)\right] \nonumber
\end{align}
With four lattice sites in the vortex lattice unit-cell, the BdG Hamiltonian in momentum space becomes a $32\times32$ matrix which can be diagonalized in the reduced 1BZ.

\section{Results}\label{secResults}

\subsection{Topological insulator normal state}\label{secNormal}

We first analyzed the non-interacting Hamiltonian (\ref{H0}) without superconductivity. We introduced an edge at $x=0$, no longer leaving $k_x$ as a good quantum number. Fourier transforming $H_0$ and analyzing the energy spectrum, we found that a topologically insulating phase can be achieved whenever the $\Delta_{X}$ and $\Delta_{\Gamma}$ terms obey the relation $ \Delta_{X}/\Delta_{\Gamma} \leq -2$. Note that $\Delta_{\textrm{M}}\neq 0$ is needed to gap out the Dirac quasiparticles at the M-point of the microscopic 1BZ and produce a surface spectrum with three Fermi pockets at $\Gamma$ and two X points, similar to the one obtained in Kondo topological insulators \cite{Kang2013, Lu2013b, Nikolic2014b}. The hopping term $t$ and the lattice site spacing $a$ are set to 1. The energy spectrum is presented in Figure \ref{TISpectra}. The emerging edge states can be seen crossing $E=0$ at the $\Gamma$ point.

After determining that our model contains a TI phase, we introduce the spin-triplet superconducting vortex lattice via (\ref{HI}). In the following analysis, we typically set the superconducting order parameter to $\bar{\eta} = t = 1$ and vary the strength $\alpha$ of the spin-orbit coupling to observe the gradual evolution of the quasiparticle spectrum. At other times, we vary $\bar{\eta}$. The proper value of $\bar{\eta}$ is ultimately determined by minimizing the energy of the superconducting state for the given chemical potential $\mu$, spin-orbit coupling $\alpha$ and interaction couplings $U_t, U_{t0}$ in (\ref{HI}) -- we do not pursue this energy minimization here because it was done before \cite{Nikolic2014a}, and because our present interest is only the qualitative nature of the quasiparticle excitations.

\subsection{The protection of Dirac points}\label{secProtection}

\begin{figure}[!t]
\includegraphics[width=0.45\textwidth]{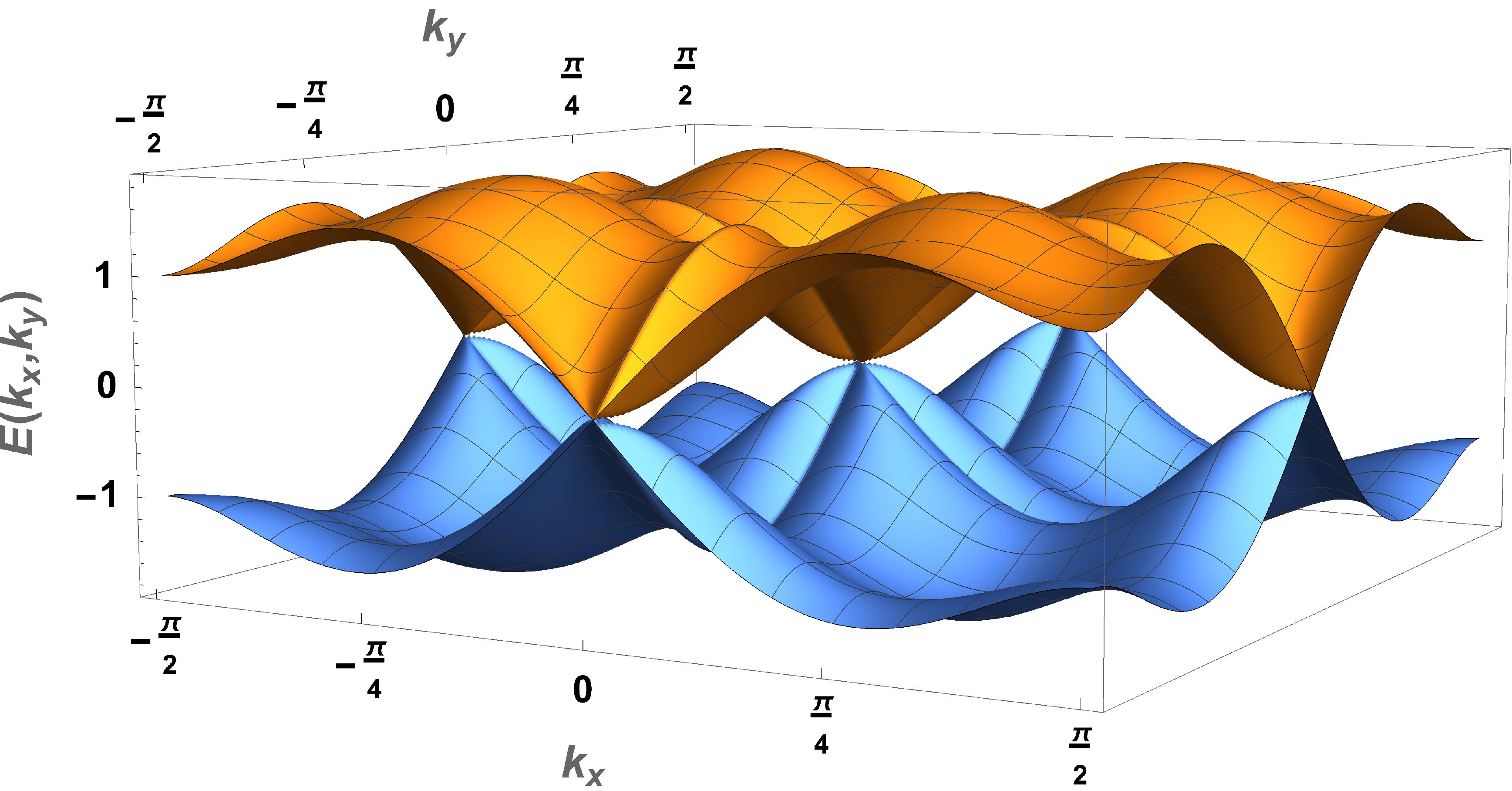}
\caption{\label{3DPlot}The lowest energy bands in the reduced Brillouin zone for the extreme spin-orbit coupling $\alpha = \frac{\pi}{2}$. The superconducting order parameter is set to $\eta = 1$. The bands are two-fold degenerate, with protected Dirac points appearing at the $\Gamma$ and X locations.}
\end{figure}

Diagonalizing the Hamiltonian produces a band structure with multiple Dirac points, including ``accidental'' and symmetry-protected ones. The extreme case $\alpha=\pi/2$ is shown in Figure \ref{3DPlot}. The inclusion of sufficiently strong superconductivity reintroduces the Dirac points at the high-symmetry wavevectors of the reduced 1BZ even when all $\Delta_{ij}$ terms are non-zero and the parent normal state is fully gapped. Additional ``accidental'' Dirac points appear in the zone interior for a range of generic $\bar{\eta}, \alpha$ values, and migrate when the model parameters change gradually. 

Once the unphysical doubling of the BdG spectrum due to the Nambu representation is taken into account, all physical bands of our model remain two-fold degenerate for any spin-orbit coupling strength $\alpha$. This degeneracy is protected by the combined presence of the time-reversal symmetry $\mathcal{T}$ and the ``film surface exchange'' symmetry under $\mathcal{I}^z=\tau^x \sigma^z$. As a consequence, every Dirac node is two-fold degenerate. When a small symmetry-breaking $\sigma^z$ or $\tau^z$ perturbation is introduced, the previously degenerate nodes split and reveal their chiral spin-momentum-locking texture. This texture is characterized by an integer winding number which distinguishes nodes from antinodes. The winding number in each band is a topologically protected ``charge'' located at the node wavevector. The evolution of these ``charges'' with model parameters in one band is mirrored in the other degenerate partner band due to the $\mathcal{T}$ symmetry.

We can also scrutinize the effect of other perturbations which respect the time-reversal, lattice translation and spatial point-group symmetries of the considered superconducting state. Let us first observe that all Dirac points which are possible in the normal state are bound to the high-symmetry wavevectors of the microscopic 1BZ, so they end up living at the $\Gamma$ point of the folded zone when the superconducting state spontaneously breaks the translation symmetry. Consequently, any Dirac point found in the superconducting state away from the $\Gamma$ point is necessarily created by the vorticity of the superconducting state itself. One mechanism of replacing a Dirac point with a gap involves connecting a Dirac particle-like state of positive energy $E=v\delta k$ with a hole-like Dirac state of negative energy $E=-v\delta k$, as in the effective Hamiltonian:
\begin{equation}
\delta H = \left( \begin{array}{cc} v\delta k & \delta \\ \delta & -v\delta k \end{array} \right) \ .                                                                                                                                                                                                                                                                                                                                                                                                                                                                                                                                                                                                                                                                                                                                                                                                        \end{equation}
Since these are the Bogoliubov quasiparticle states of a superconductor, the perturbation $\delta$ will generally include some kind of Cooper pairing. In order for it to not break the time-reversal symmetry, it must be either a singlet pairing, or a kind of triplet pairing  (\ref{SCParam}) that our order parameter realizes. Both of these perturbations are energetically discouraged by the spontaneous symmetry breaking that takes place in our model \cite{Nikolic2014a}.

Going beyond pairing perturbations, a density perturbation can be generated by $\delta\neq 0$, but it amounts to a benign shift of the chemical potential. A small perturbation $\lambda \tau^z$ in (\ref{H0}), which would arise in a gated and biased sample, lifts the two-fold band degeneracy and shifts any two accidental degenerate Dirac cones at the same wavevector in opposite energy directions, without initially creating a gap. The two ``vertically`` separated Dirac cones intersect at zero energy on a small ring in momentum space, and the quadratic band touching created by the accidental Dirac point merger survives at a shifted value of the spin-orbit coupling $\alpha$. Any modification of the spin-orbit coupling that introduces an SU(2) gauge field (\ref{Gauge}) on further-neighbor site pairs also generally creates additional Dirac points in the microscopic 1BZ (a new qualitative feature that we rule out). Other small further-neighbor hopping and $\Delta$ perturbations can only gradually move the Dirac points according to our calculations. Therefore, a natural symmetry-based protection mechanism is in place to preserve the gapless tunable Dirac nodes of the superconducting state.

Translation symmetry-breaking perturbations are generally able to gap-out pairs of Dirac points which are separated in momentum space by the perturbation's characteristic wavevector. Disorder is such a perturbation in principle, but it will have a limited effect if it respects the lattice translation symmetries on average.

\subsection{The migration of ``accidental'' Dirac points}\label{secMigration}

The quasiparticle spectrum exhibits an intricate evolution as a function of the spin-orbit coupling strength $\alpha$ and superconducting amplitude $\bar{\eta}$. In the following, we keep $\bar{\eta}=1$ and consider variation of $\alpha$. The extreme value $\alpha=\pi/2$ corresponds to the case with a special symmetry on the square lattice, as discussed in Ref.\cite{Nikolic2014a}. The model is periodic under a transformation of $\alpha \rightarrow \alpha + 2\pi$. The transformation $\alpha \rightarrow \alpha+ \pi$ is equivalent to a change in the sign of the hopping term $t$. The inversion symmetry of the model (when both spin and lattice transformations are included) yields the energy spectrum which is invariant under a sign change of the spin-orbit coupling, $\alpha \rightarrow -\alpha$. We neglect the high energy states in the spectrum and consider only the two bands with Dirac points that lay closest to the Fermi energy $\mu=0$. Some Dirac points move through momentum space as $\alpha$ varies, but the Dirac points located at ($0,0$), ($\pi/2, 0$), ($0, \pi/2$) and ($\pi/2, \pi/2$) in the reduced Brillouin zone are protected by the point-group symmetries of the superconducting state and immobile regardless of the spin-orbit coupling value.

\begin{figure*}[t]
\centering
    \subfigure[{}]{
      \includegraphics[width=0.45\textwidth]{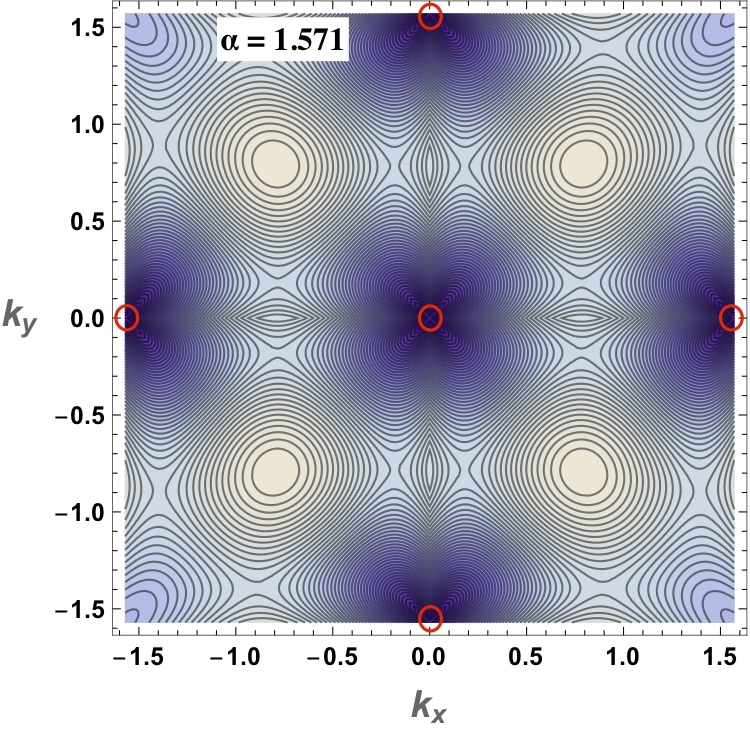}}
    \subfigure[{}]{
       \includegraphics[width=0.45\textwidth]{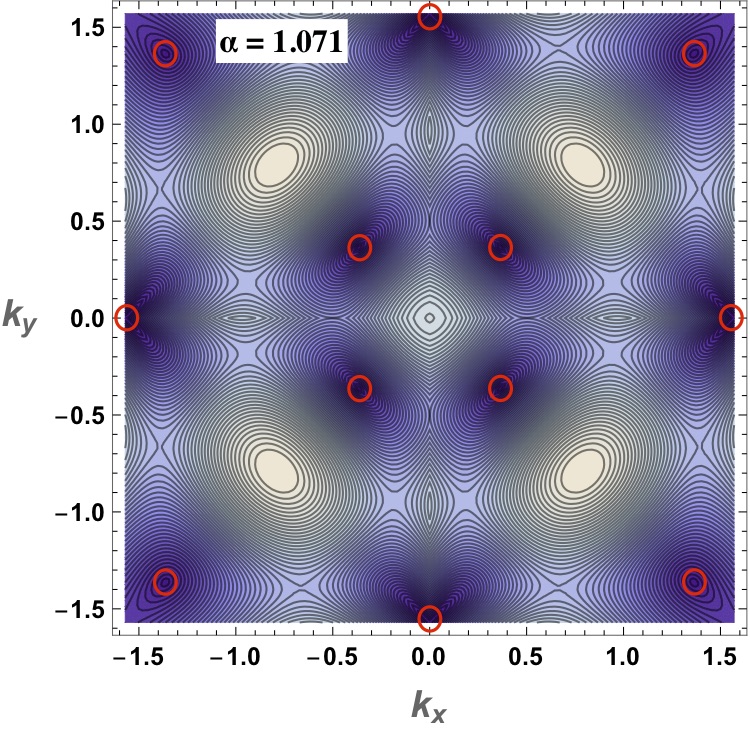}}
    \subfigure[{}]{
       \includegraphics[width=0.45\textwidth]{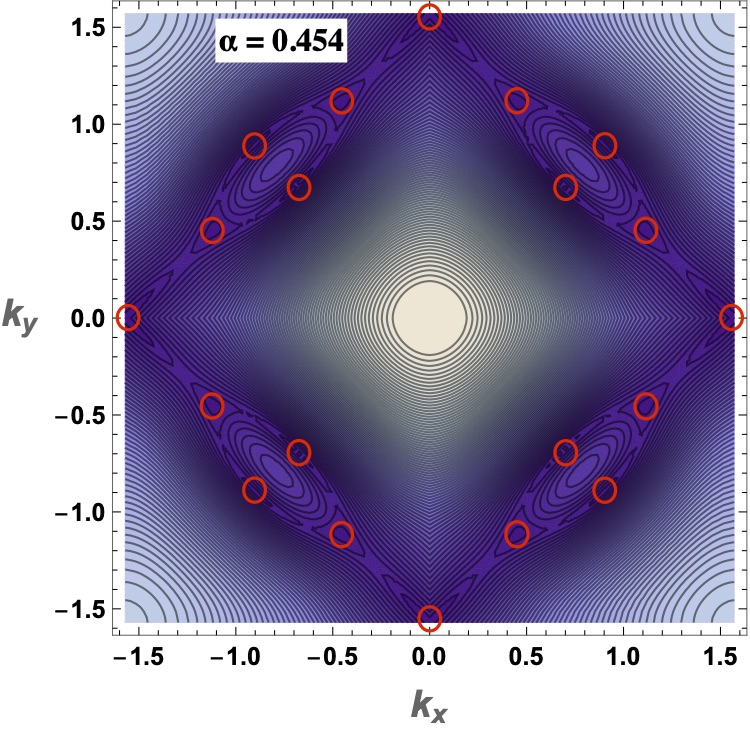}}
    \subfigure[{}]{
       \includegraphics[width=0.45\textwidth]{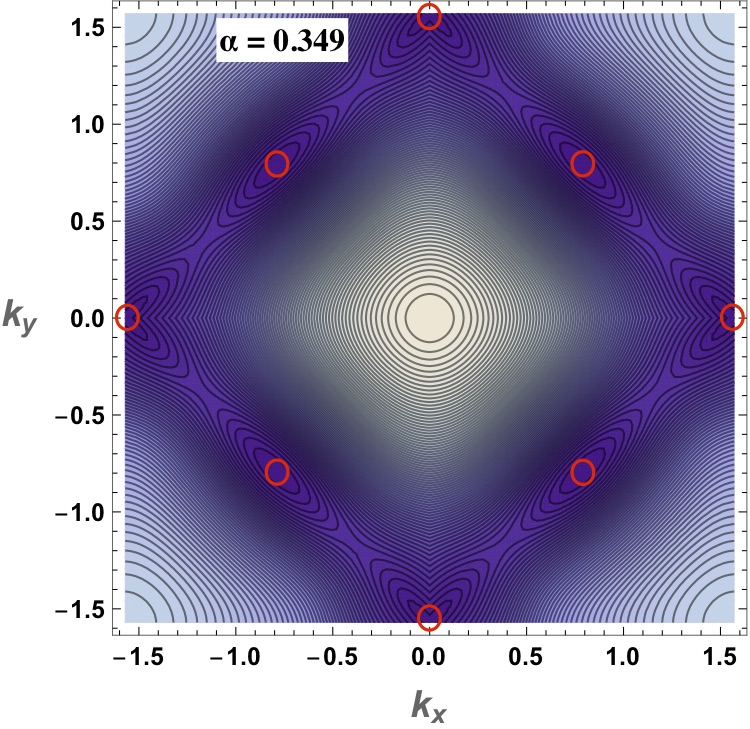}}
    \caption{Quasiparticle energy contour plots in the reduced first Brillouin zone. In the panels (a)-(c), the red circles indicate the locations of zero-energy Dirac points and their migration as $\alpha$ is varied from $\alpha = 1.571$ to $\alpha = 0.349$. In the last panel (d), a quadruplet of Dirac points has merged to temporarily produce a zero-energy quadratic band touching node represented by the red circle; this node finally gaps out at $\alpha < 0.349$.}
    \label{CPlots}
\end{figure*}

\begin{figure*}[t]
\centering
    \subfigure[{}]{
      \includegraphics[width=0.45\textwidth]{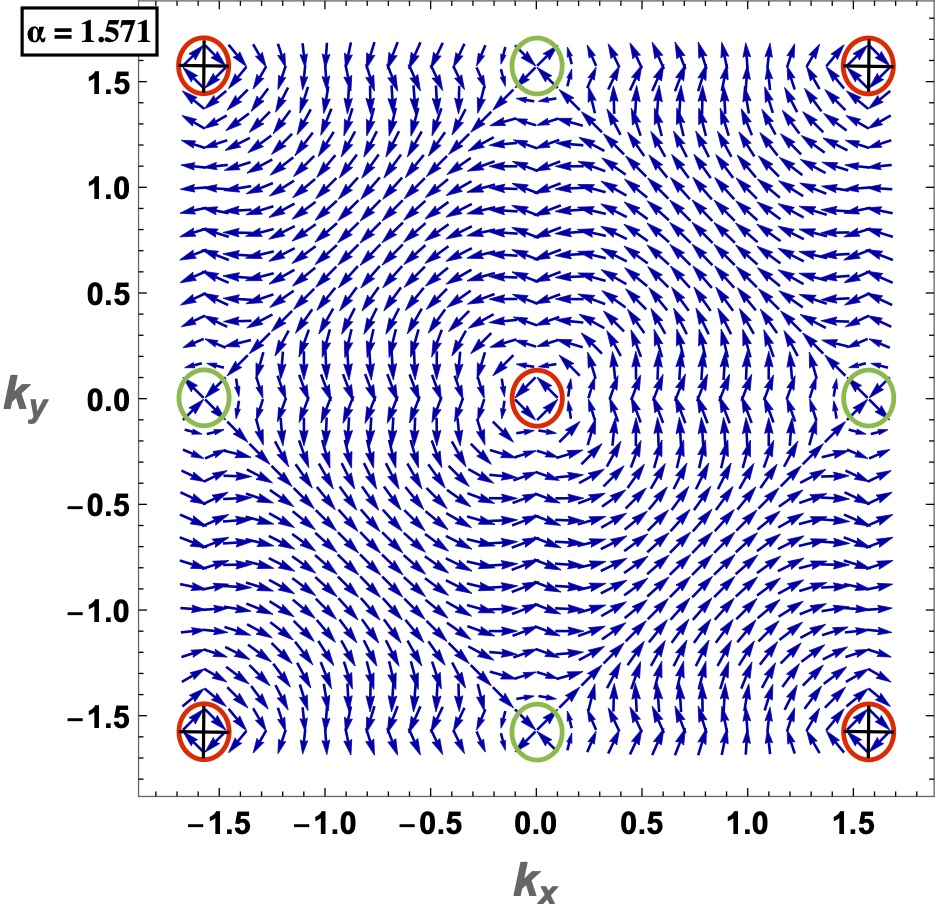}}
    \subfigure[{}]{
       \includegraphics[width=0.45\textwidth]{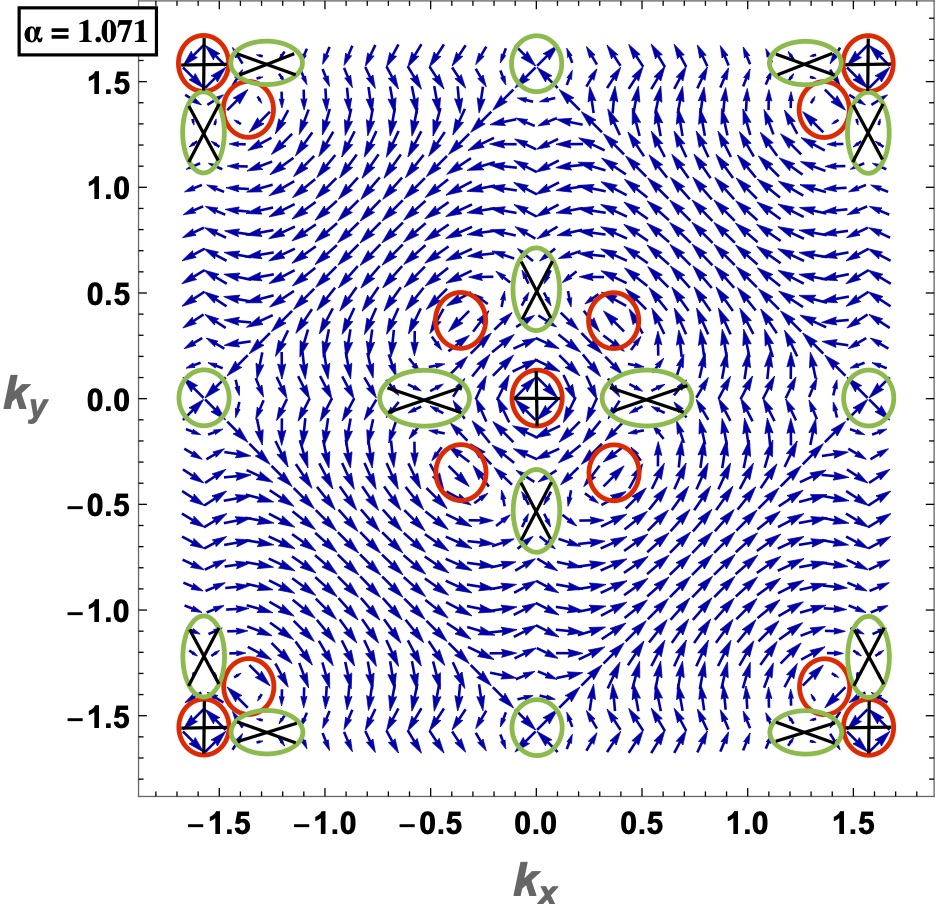}}
    \subfigure[{}]{
       \includegraphics[width=0.45\textwidth]{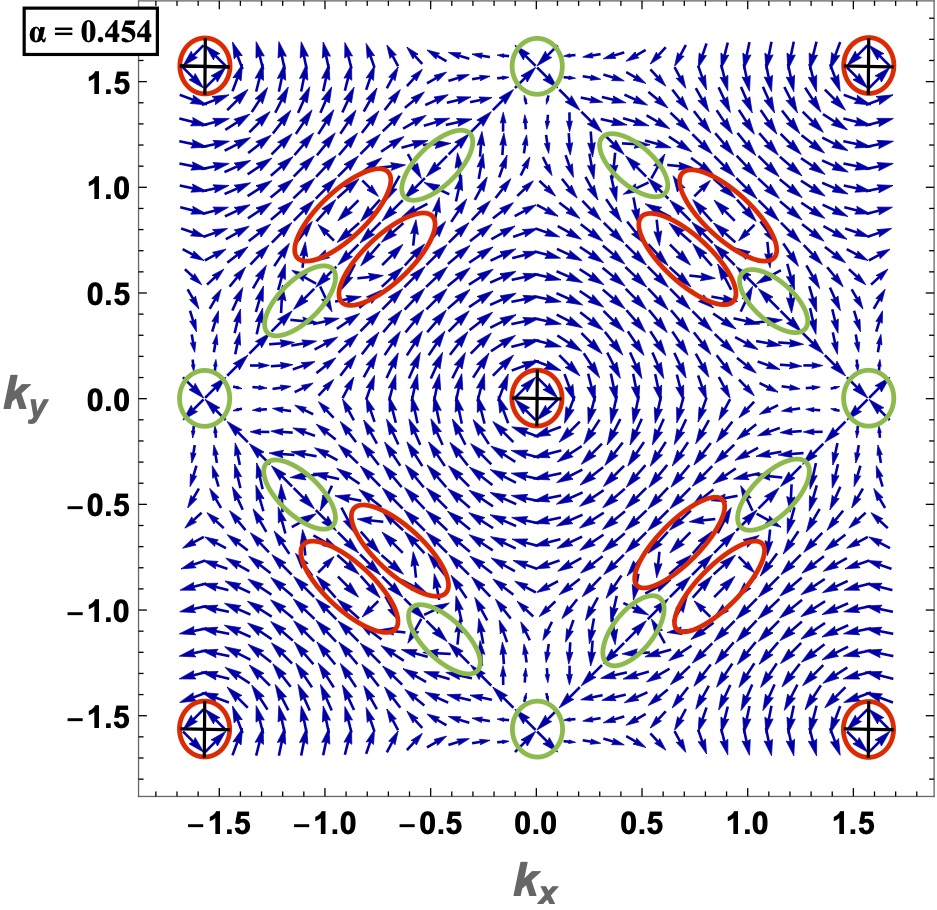}}  
    \subfigure[{}]{
       \includegraphics[width=0.45\textwidth]{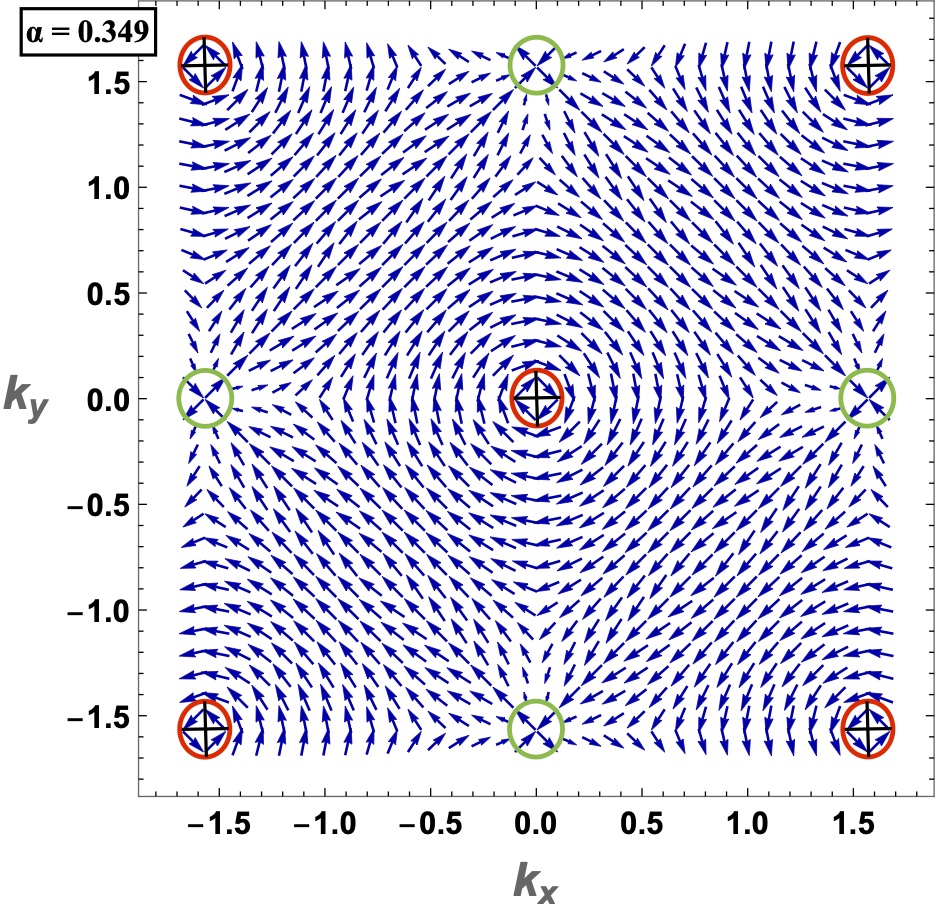}}
    \caption{Spin-momentum locking texture of the lowest-energy band in the reduced first Brillouin zone, obtained when the two-fold band degeneracy is formally lifted by an infinitesimal $\sigma^z$ or $\tau^z$ perturbation. The red and green ovals indicate vortices (nodes) and antivortices (antinodes) respectively. In comparison to Fig.\ref{CPlots}, note that the spin texture singularities are not necessarily pinned to zero energy; they may appear as Dirac points at which the shown band touches a higher energy band (this is denoted with a cross inside the oval).}
    \label{STexts}
\end{figure*}

To understand how the spin-orbit coupling affects the Dirac points, consider gradually reducing $\alpha$ from $\pi/2$ to zero. The ensuing evolution of zero-energy Dirac nodes is shown in Figure \ref{CPlots}, and the corresponding evolution of the spin-momentum locking texture is shown in Figure \ref{STexts}.

\begin{figure}[t]
    \centering
    \includegraphics[width=0.48\textwidth]{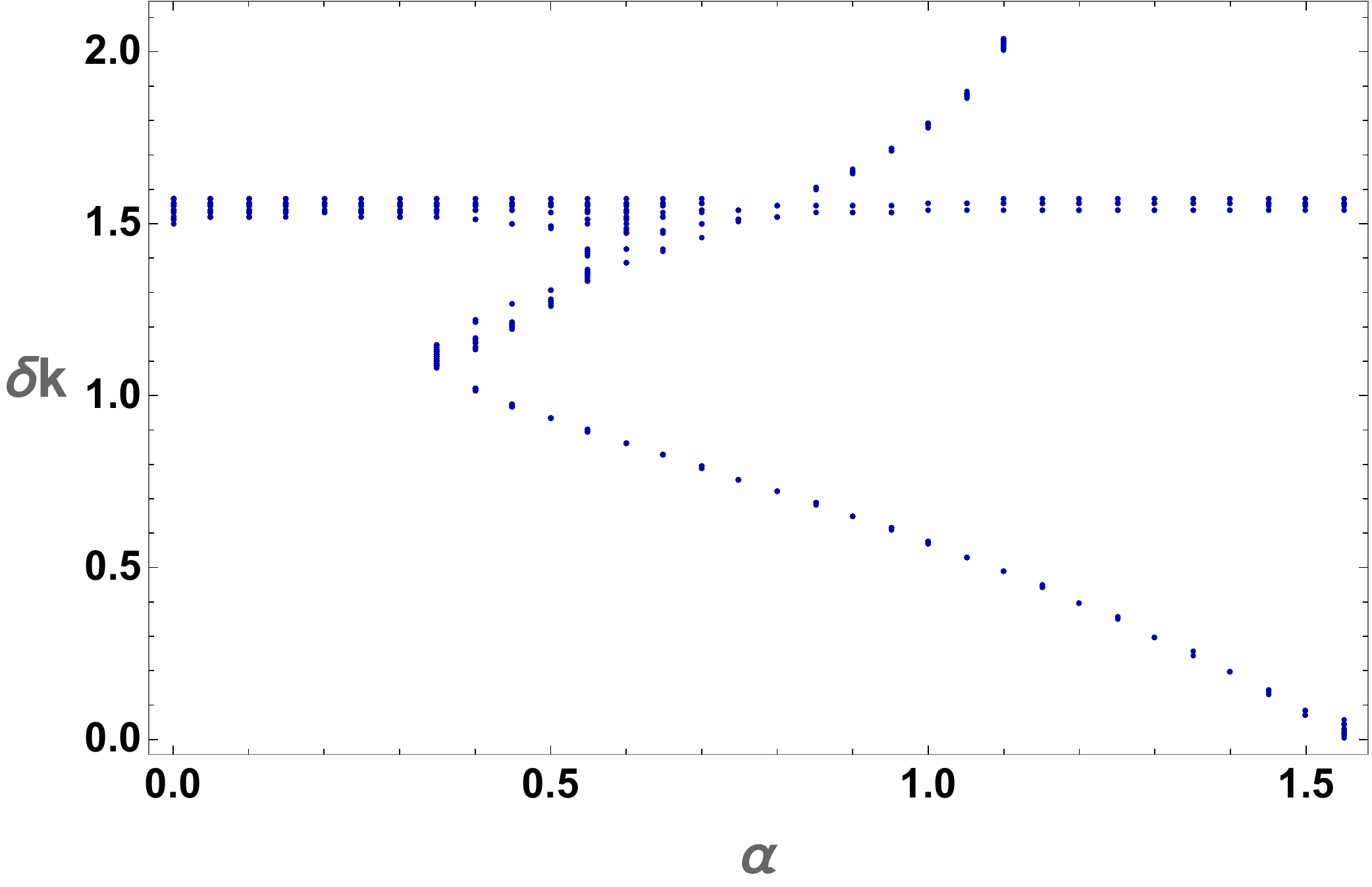}
    \caption{The locations of zero-energy Dirac nodes in the reduced first Brillouin zone as the spin orbit coupling $\alpha$ is varied ($\bar{\eta}=t=1$). Here, $\delta k =\sqrt{k_{x}^2 +  k_{y}^2}$, where  $k_{x}$ and $k_{y}$, are the $x$-component and $y$-component, respectively, of the wavevector within the first Brillouin zone where the Dirac nodes are found.}
    \label{AlphaVdk}
\end{figure}

When $\alpha \lesssim\pi/2$, four zero-energy Dirac nodes and four antinodes emerge from the $\Gamma$ point. With $\alpha$ decreasing, these nodes move apart toward the 1BZ corners along the zone diagonals and the antinodes move apart along the horizontal and vertical directions toward the X points. At $\alpha=1.126$, new Dirac points emerge from the 1BZ corners (M points) and begin migrating along the zone diagonals toward the $\Gamma$ point, accompanied by the antinodes that migrate toward the X points along the zone boundary. The four antinodes approaching each X point ``bounce off'' from the X point at $\alpha\approx 0.6$ and continue migrating along the X-X lines toward the quadrant centers as $\alpha$ is further reduced. A pair of nodes moving away from the $\Gamma$, M points and a pair of ``bounced'' antinodes moving away from the X points meet at ${\bf k}=(\pm \pi/4, \pm \pi/4)$ when $\alpha=0.349$. This coalescence of four Dirac points is found to temporarily produce a zero-energy quadratic band touching at ${\bf k}=(\pm \pi/4, \pm \pi/4)$. Going on to $\alpha<0.349$, the fused nodes are annihilated and the local spectrum becomes gapped. Figure \ref{AlphaVdk} summarizes the migration of the Dirac nodes with the varying spin-orbit coupling.

The node-antinode distinction is formally introduced only when the two-fold band degeneracy is lifted by at least an infinitesimal symmetry-breaking perturbation, $\sigma^z$ or $\tau^z$. Such a perturbation splits the band degeneracy and reveals a definite node helicity in every band. Figure \ref{STexts} shows the evolution of the spin-momentum locking texture and local helicity. Every Dirac point is a unit vortex or antivortex of the spin texture vector field, and the total topological charge of all such singularities in the 1BZ is zero due to the periodic boundary conditions of the momentum space. The spin texture provides an important means for the bookkeeping of all nodes and reveals the nature of their creation or annihilation. Specifically, we observe that the Dirac points are always created and annihilated in quadruplets of two nodes and two antinodes. These types of quadruplets are not only promoted by the symmetries of the square lattice, but also turn out to be a hallmark of the Rashba-type spin-orbit coupling even in the continuum limit. In real space, quadruplets are the \emph{elementary} neutral vortex clusters of triplet superconductors shaped by the Rashba spin-orbit coupling \cite{Nikolic2014}. The superconducting vortex lattice considered here is a stable tiling of vortex quadruplets. Perhaps not surprisingly, the coalescence of Dirac nodes is a momentum-space incarnation of the quadruplet annihilation -- also responsible for the quadratic band touching; the simpler single node-antinode annihilation yields a parabolic dispersion only \emph{after} the gap opens. Note that the nodes shown in the spin texture are not required to live at zero energy because Dirac points can also connect two bands at a higher energy. For example, even though $\Gamma$ point breaks up into the four diagonally-moving Dirac points, there is still a vortex, as seen from the spin texture, that persists at the $\Gamma$ point but at a higher energy (touching another band).

\subsection{Quadratic band touching}\label{secQuadratic}

\begin{figure}[t]
\centering
    \subfigure[{}]{
      \includegraphics[width=0.48\textwidth]{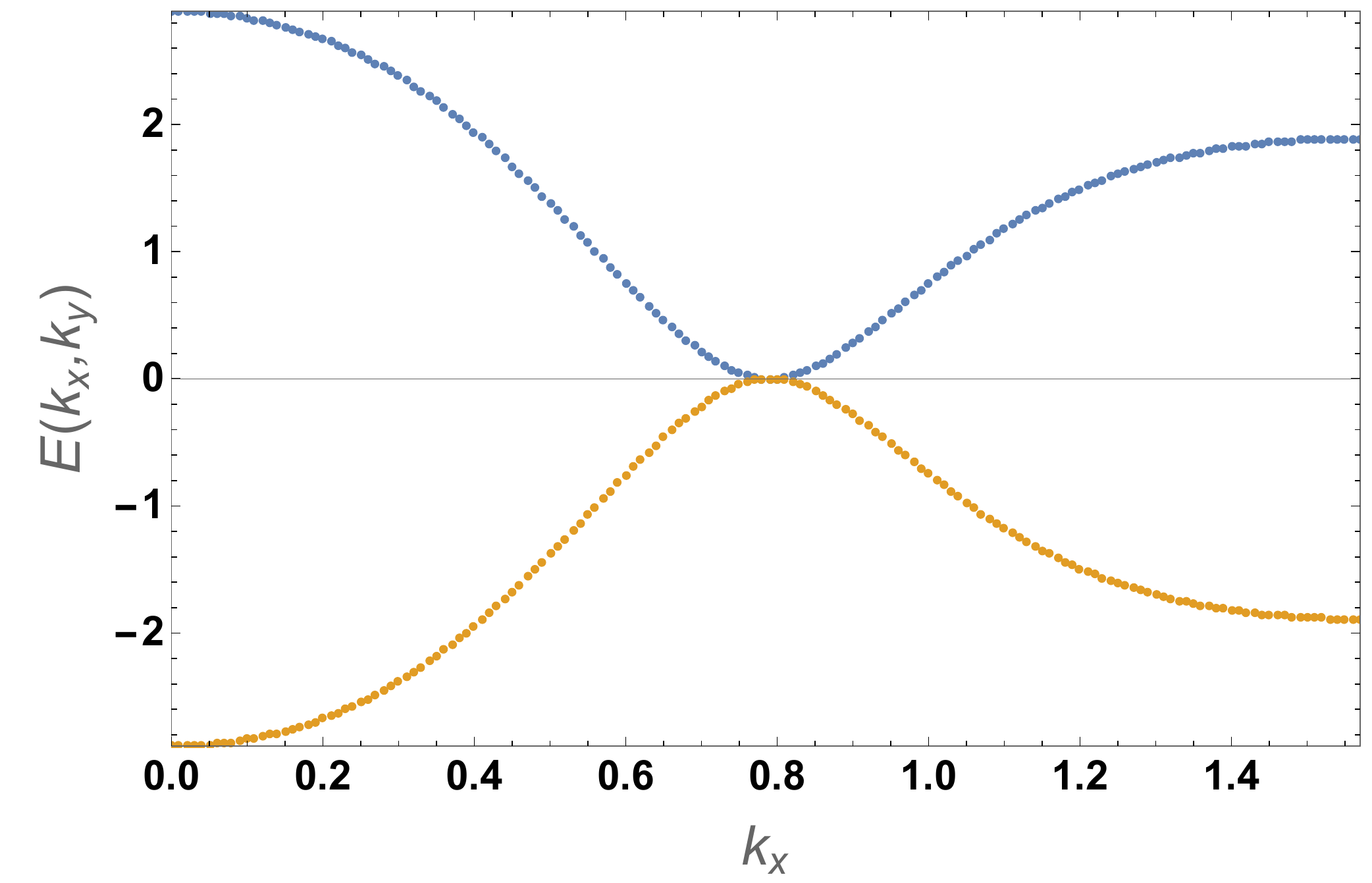}}
    \subfigure[{}]{
       \includegraphics[width=0.48\textwidth]{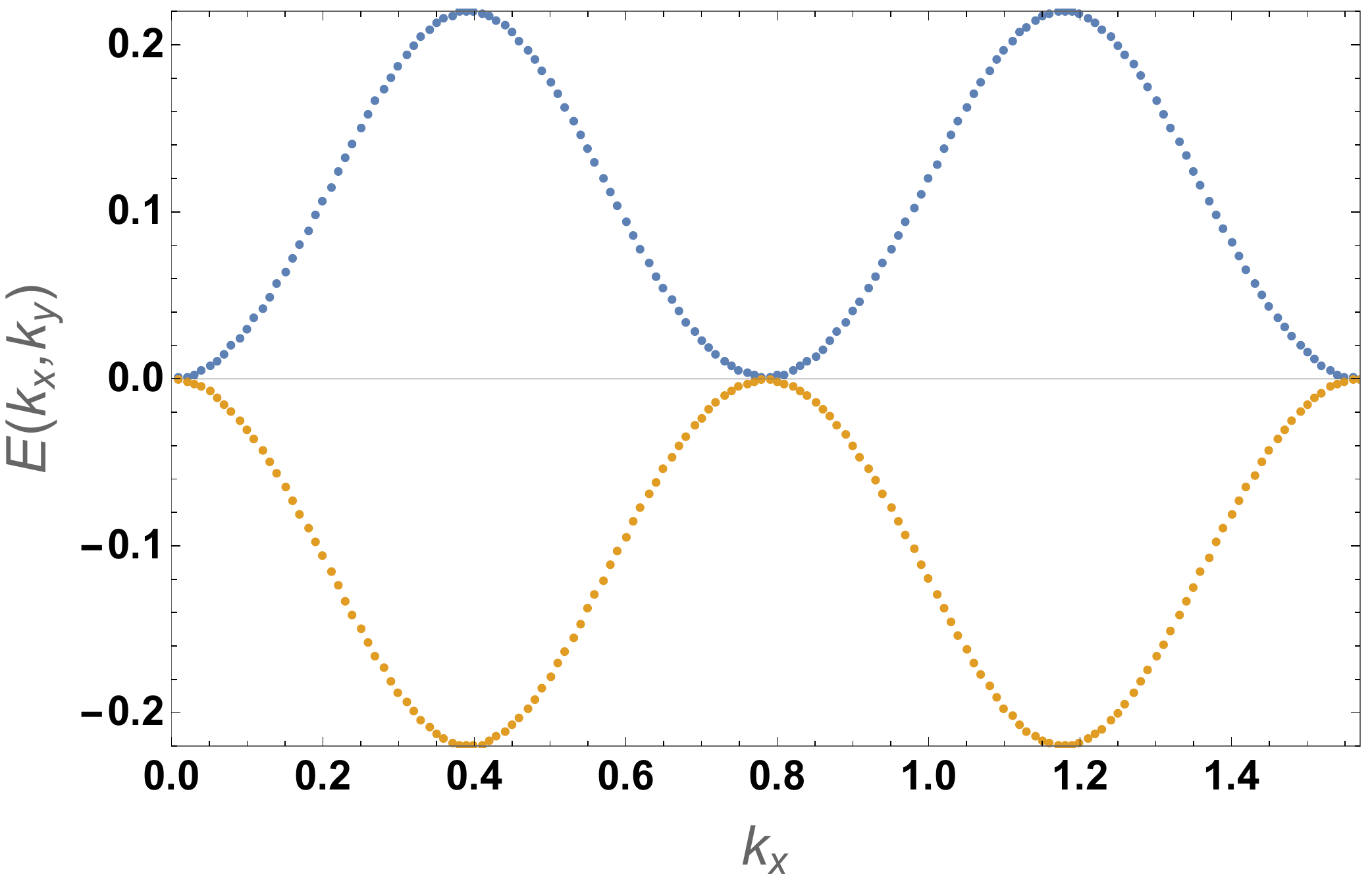}}
    \caption{Quadratic band-touching dispersion in the quasiparticle energy spectrum at $\alpha = 0.349$: (a) along the $\Gamma$-M line, (b) along the X-X line. The quadratic node is visible at $k_x=\pi/4$, and corresponds to the vanishing net topological charge of the merging Dirac nodes and antinodes. The nodal spectrum near $k_x=0$ and $k_x=\pi/2$ is actually linear, Dirac-like, even though it is not obvious from this figure. This is driven by the unit topological charge pinned to the high-symmetry wavevectors (see Fig.\ref{STexts}), and indeed a small linear $v$ term in the local dispersion $E \approx vk+uk^2$ is found there.}
    \label{ELines}
\end{figure}

The emergence of quadratic band touching from the coalescence of Dirac point quadruplets at $(\pm\pi/4, \pm\pi/4)$ is unusual, so we characterize it in several different ways. Figure \ref{ELines} shows the band structure slices through the coalescence point $(\pi/4, \pi/4)$ in two characteristic directions; the quadratic dispersion ($k_x=\pi/4$) is evident.

\begin{figure*}[t]
\centering
    \subfigure[{}]{
      \includegraphics[width=0.45\textwidth]{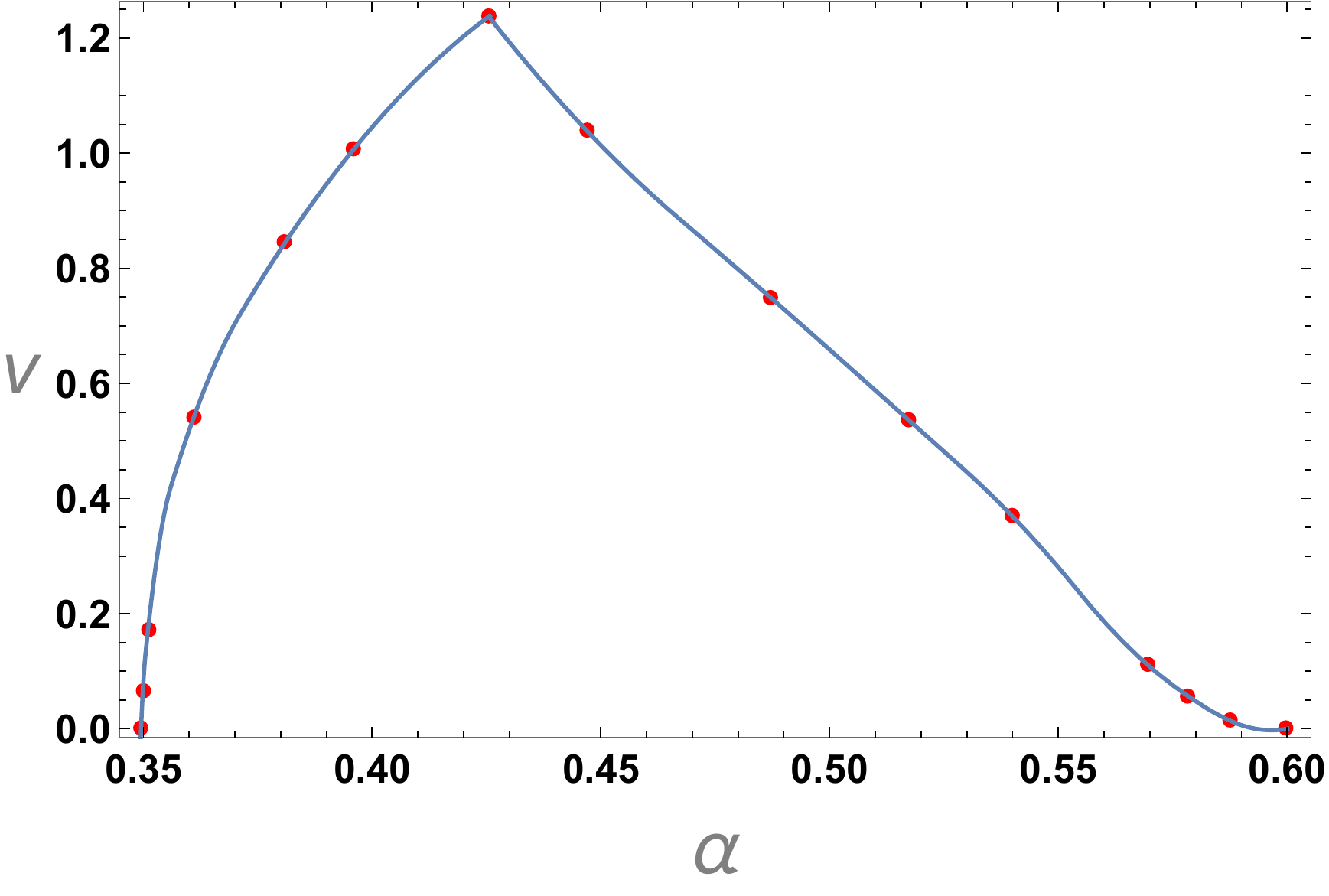}}
    \subfigure[{}]{
       \includegraphics[width=0.45\textwidth]{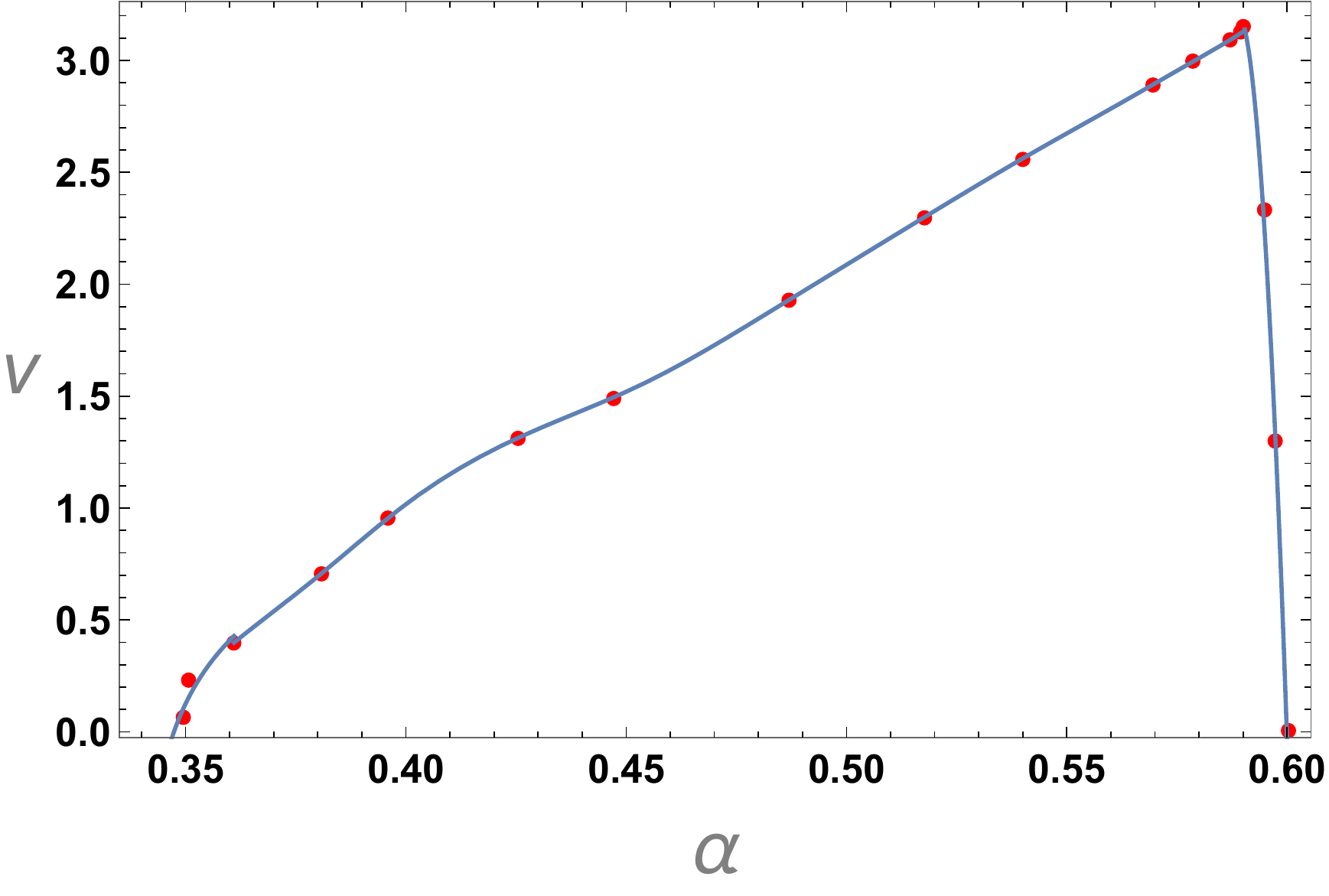}}
    \subfigure[{}]{
       \includegraphics[width=0.45\textwidth]{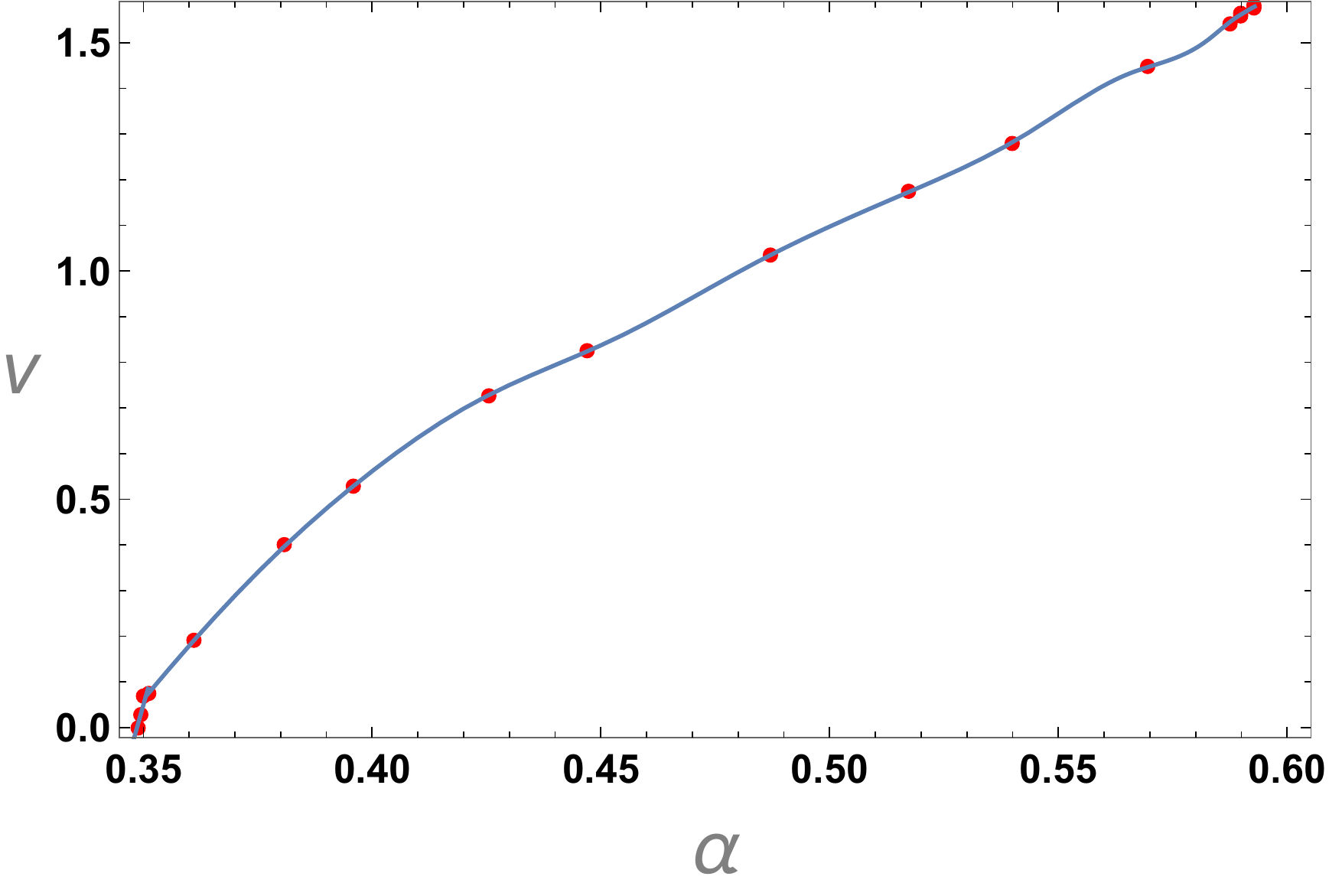}}
    \subfigure[{}]{
       \includegraphics[width=0.45\textwidth]{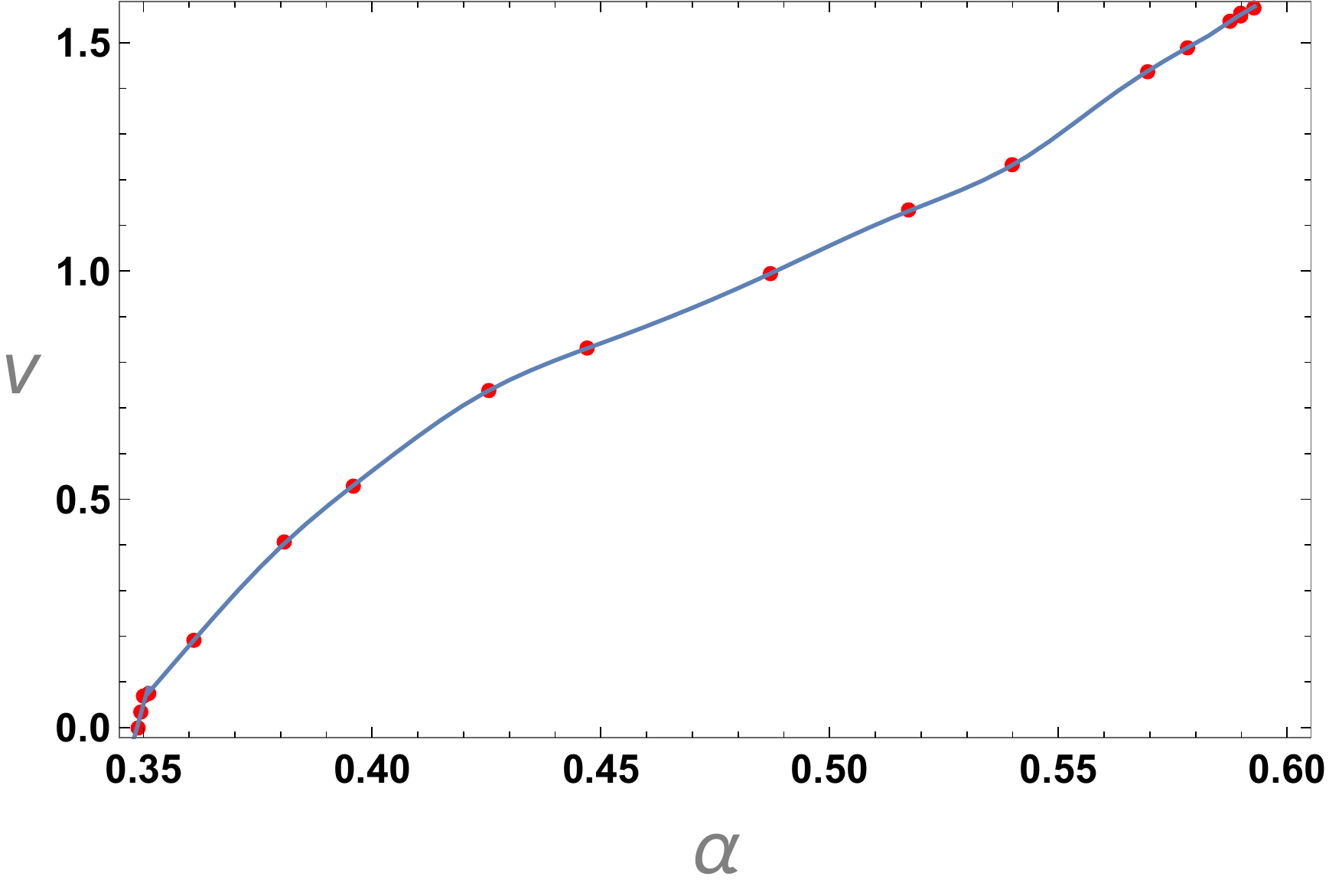}}
    \caption{The anisotropic Fermi velocities measured at the ``accidental'' Dirac antinodes emerging from the X points as they migrate with the variations of spin-orbit coupling $\alpha$. The plots show the Fermi velocities in four different directions: (a) along the line connecting the two X points, (b) parallel to the $\Gamma$-M line,  (c) vertical and (d) horizontal. The Fermi velocity approaches zero in all directions as the antinode approaches the merger annihilation point at $\alpha=0.349$. The value $\alpha\approx0.6$ corresponds to the antinode's proximity to the X point.}
    \label{FermiPlots}
\end{figure*}

To further ensure that the node coalescence results in a quadratic band touching, we tracked the Fermi velocity of the Dirac nodes in four separate momentum-space directions during the Dirac node migration. This is shown in Figure  \ref{FermiPlots}. The Fermi velocity was determined for the  nodes in the first quadrant of the reduced Brillouin zone, but the symmetry of the system dictates that these velocities be the same for the nodes in the remaining three quadrants. When the spin-orbit coupling is reduced, the Fermi velocities in the four measured directions approach zero as the node wavevectors approach ${\bf k}=(\pm \pi/4, \pm \pi/4)$. This indicates that the band at this wavevector is quadratic in all directions. The nodal spectrum at the $\Gamma$ and M points was found to be linear and isotropic at low energies for all $\alpha$ values, even when new Dirac nodes emerge. This is related to the fact that the topological charge is always non-zero at these points in the 1BZ.

Quadratic features of the nodal spectrum are also seen at the X-points of the 1BZ when $\alpha=0.6$. This is the instant at which four Dirac anti-nodes bounce off from each X-point and switch their migration from the zone-boundary to the ``diagonal'' directions. Figure \ref{FermiPlots}(a,b) shows that the Fermi velocity becomes zero along the ``diagonal'' directions at $\alpha=0.6$ for the Dirac anti-nodes emerging from the X points. However, the X point does not host a proper quadratic band-touching node. The panels (c,d) of Figure \ref{FermiPlots} indicate that a linear nodal dispersion remains in the horizontal/vertical directions, so these coalescence nodes have the most unusual structure. This is correlated with the fact that a fixed antinode persists at the X points throughout the coalescence and bounce-off process.

\begin{figure}
    \centering
    \includegraphics[width=0.48\textwidth]{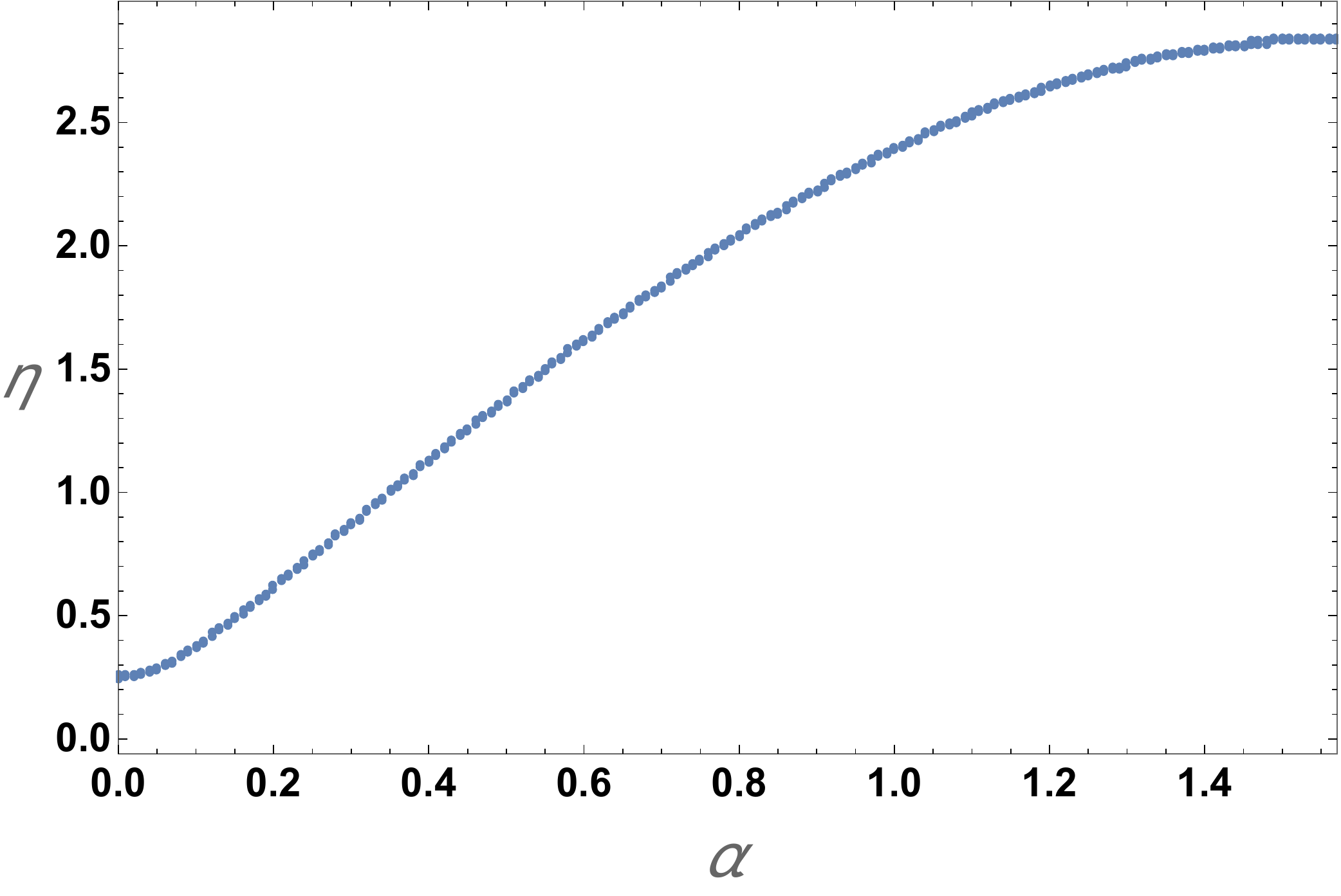}
    \caption{The function $\eta\left(\alpha\right)$ shows the values of the superconducting order parameter strength $\bar{\eta}$ for any given spin-orbit coupling $\alpha$ at which the Dirac point quadruplets merge into quadratic band-touching nodes at the wavevectors ${\bf k}=(\pm \pi/4, \pm \pi/4)$. The function is symmetric with respect to $\alpha=\pi/2$.}
    \label{Eta_V_Alpha}
\end{figure}

The coalescence of the Dirac nodes is determined by both the spin-orbit coupling $\alpha$ and the superconducting order parameter amplitude $\eta$. The variations of either can produce a quadratic node at ${\bf k}=(\pm \pi/4, \pm \pi/4)$. Figure  \ref{Eta_V_Alpha} shows the relationship between the spin-orbit coupling and the superconducting order parameter which yields these quadratic nodes. In a laboratory, one would indirectly control the order parameter amplitude $\eta$ by varying the gate voltage, while the spin-orbit coupling remains microscopically fixed. The migration of the Dirac nodes and their creation or annihilation can be, therefore, driven by the gate voltage as long as the underlying vortex lattice phase is stable. Indeed, the vortex lattice of spin currents has a broad range of stability within a simple representative model \cite{Nikolic2014a}; realistic extensions of that model would affect the range of stability, and the conclusions of our present study apply only within that range. The main new insight from this study is that the spin current vortex lattice phase can host additional topological phase transitions at which Dirac nodes are created or destroyed, and that it remains generally stable across these transitions.

\section{Experimental signatures}\label{secExperiments}

A two-dimensional triplet superconductor hosting a vortex lattice of spin currents is not a state that can be easily identified in an experiment. Being a paramagnet, this state would not produce any interesting magnetic response; magnetic susceptibility is found to be thermally activated. Superconductivity in the TI film can be readily detected in transport measurements as a rapid increase of conductivity below a critical temperature, which can be suppressed with a gate voltage (when the electrons are pushed out out of the film). However, the presence of spin current loops is not directly observable with any known probe. In order to see the broken lattice symmetry with scanning tunneling microscopy (STM), a way to locally distinguish spin current vortices and antivortices must be devised, perhaps by tunneling spin-polarized electrons in a biased orbital angular momentum state.

Observing the signatures of nodal quasiparticles is evidently the best way to recognize the appearance of this state in a physical system. The presence of Dirac quasiparticles which robustly persist under the variations of a gate voltage would also be a strong indicator of the non-trivial topological character of the superconducting state.

There are a number of surface probes which can directly characterize the quasiparticle spectrum. ARPES with a sufficient resolution can visualize the quasiparticle spectrum $E({\bf k})$ and identify the presence and character of nodes. STM could detect the Dirac spectrum by measuring the density of states as a function of energy (tip bias). Even Raman scattering can identify Dirac electrons \cite{Wang2008b, Kashuba2009, Nikolic2022a}. These methods would be best suited for the alternative realization of the spin current vortex lattice on the surface of a Kondo topological insulator. As explained in the introduction, the surface states of a Kondo insulator such as SmB$_6$ carry the same degrees of freedom with a similar basic dynamics as the superconducting system we focused on. Even though the pairing can there occur in the exciton channel, the quasiparticle excitations would inherit the same nodal character. The problem with the superconducting realization is the need for a proximity effect. A TI film is placed between a superconducting and an insulating material in an heterostructure device \cite{Nikolic2011a}, so it is not directly accessible to the surface probes unless some pattern of heterostructure layers can be engineered to allow access at a grid of locations.

Anomalous temperature dependence of the specific heat is in principle a simple probe of Dirac quasiparticles in any system. Fermionic quasiparticles with a nodal dispersion $E\propto k^n$ in $d$ dimensions are expected to contribute a $T^{d/n}$ behavior to the specific heat. In $d=2$ dimensions, the $C\sim T^2$ signature of Dirac quasiparticles is exposed and cannot be confused with either an insulating or Fermi liquid behavior. Fortunately, gapless two-dimensional bosonic excitations which would also produce $C\sim T^2$ are absent: phonons cannot be confined to the crystal boundary, and the Goldstone modes of a superconducting state are gapped by the Anderson-Higgs mechanism. At special points of the phase diagram, the emerging quadratic band-touching nodes would temporarily introduce a linear $C\sim T$ temperature dependence of the specific heat ($d=2, n=2$). This also cannot be confused with the Fermi liquid behavior because it would require an unusual fine-tuning of the gate voltage across the ultrathin-film sample. In practice, all these behaviors of the TI film would need to be extracted with high sensitivity by subtracting the specific heat of the system from that measured in the normal state (with a gate voltage set to push the electrons out of the film). 

\begin{figure}[t]
    \centering
    \includegraphics[width=0.48\textwidth]{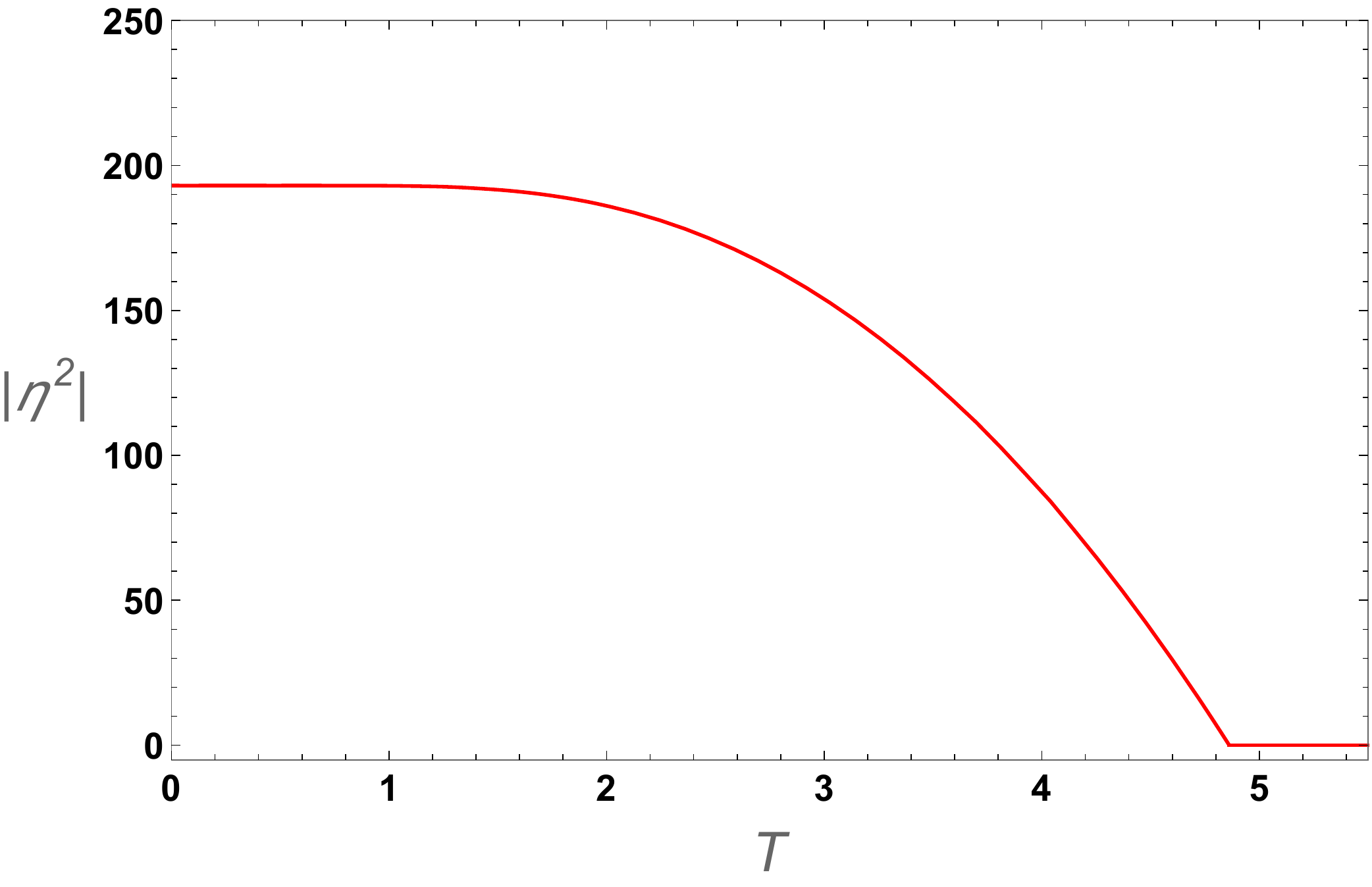}
    \caption{The magnitude of the superconducting order-parameter is computed as a function of temperature. The critical temperature of the system occurs at 4.86$t$.}
    \label{Tc}
\end{figure}

The presence of a superconducting state would also leave a thermodynamic footprint in phase transitions. The continuous U(1) symmetry cannot be spontaneously broken in two dimensions, but a Kosterlitz-Thouless transition is expected at a finite critical temperature $T_{\textrm{KT}}$. A crude mean-field upper bound $T_{\textrm{mf}} = 4.86t$ for this transition temperature is obtained numerically by computing the temperature dependence of the superconducting order parameter in our model, shown in Figure \ref{Tc}. The broken lattice translation symmetry is discrete, so another phase transition above $T_{\textrm{KT}}$ may be required to restore all symmetries. We have not studied this transition, but naively expect that the mean-field result $T_{\textrm{mf}}$ estimates its critical temperature better.

\section{Conclusions and discussion}\label{secConclusions}

We studied the quasiparticle spectrum of a prominent superconducting state that can be stabilized in ultrathin topological insulator films. We modelled the film with a tight-binding Hamiltonian on the square lattice and mathematically represented the spin-orbit coupling with an effective SU(2) gauge field. When generic phonon-mediated attractive interactions among the electrons produce superconductivity, the strong spin orbit coupling favors triplet Cooper pairs and condenses them into a state that hosts a lattice of spin-current vortices. This is similar to the formation of an Abrikosov lattice in an ordinary type-II superconductor subjected to an external U(1) magnetic field. The quasiparticle excitations in this superconducting state are found to have several unusual properties.

The spectrum exhibits Dirac points of the Bogoliubov quasiparticles pinned to zero energy, which gradually evolves as a function of the spin-orbit coupling and order parameter strengths. The Dirac nodes in momentum space carry positive or negative topological charge revealed by the vortex or antivortex configuration of the spin-momentum locking texture. When the model parameters are varied, the nodes are created and annihilated only in 2-node-2-antinode quadruplets, a motif characteristic for the spin-triplet condensates shaped by the Rashba-type spin-orbit coupling. Some of the nodes are pinned by symmetries to the high-symmetry wavevectors of the first Brillouin zone, while others migrate through the zone along straight lines. At the points of quadruplet creation or annihilation, the quasiparticle spectrum temporarily becomes a quadratic band touching -- this is a ``critical point'' separating a gapped ``phase'' from a Dirac node ``phase'' (locally in momentum space). At least in principle, these unusual nodal features make it possible to experimentally identify the spin-current superconducting state using thermodynamic probes and ARPES.

The topological superconducting state we considered has not been realized in any material yet, but it is a prominent mean-field ground state of a simple natural model that can minimalistically describe realistic systems. The original system proposed for this model was an ultrathin topological insulator film with a strong spin-orbit coupling placed in proximity to a conventional but strong superconductor. By embedding such a system in a gated heterostructure device, it becomes possible to control the presence and the strength of its superconducting order parameter through the gate voltage. We have shown here that this in turn controls the number and momentum-space locations of gapless Dirac quasiparticles whose chemical potential is strictly pinned to the node at zero energy by the particle-hole symmetry. Another physical system in which this kind of a state might be possible to realize is the surface of a Kondo topological insulator. However, an exciton rather than Cooper pair condensate would here be more natural, and the condensation would need to take advantage of the electron scattering between different helical Fermi pockets (on the surface-state Dirac cones) facilitated by Coulomb interactions. Gate control would also be harder to imagine because this system is fundamentally three-dimensional.

Despite being hard to realize a triplet topological superconductor with a vortex lattice, its properties are very appealing in comparison to the widely studied Weyl semimetals. In many regards, the quasiparticles of the state we study are a tunable two-dimensional version of a Weyl semimetal. But, in contrast to Weyl semimetals, the very desirable pinning of the chemical potential to the node energy is inevitable instead of being elusive. The ensuing true topological semimetal properties would be shunted by the superconductivity in the charge and spin transport, but not at all in thermodynamics or spectroscopy. Interaction effects involving the nodal quasiparticles would surely be interesting to study in two dimensions. Especially interesting are the quasiparticles that can be tuned to their quadratic band touching nodes -- their elevated density of states at zero energy makes them prone to new instabilities.

\section{Acknowledgements}

This work was supported by the Quantum Science and Engineering Center at George Mason University (GMU), and also performed at the GMU Center for Quantum Science. R.M. was supported by the Special Physics and Astronomy Graduate Research Assistantship, and Summer Research Fellowships at GMU. A part of this project was supported by resources provided by the Office of Research Computing at GMU and funded in part by grants from the National Science Foundation (Awards Number 1625039 and 2018631).


\begin{thebibliography}{59}%
\makeatletter
\providecommand \@ifxundefined [1]{%
 \@ifx{#1\undefined}
}%
\providecommand \@ifnum [1]{%
 \ifnum #1\expandafter \@firstoftwo
 \else \expandafter \@secondoftwo
 \fi
}%
\providecommand \@ifx [1]{%
 \ifx #1\expandafter \@firstoftwo
 \else \expandafter \@secondoftwo
 \fi
}%
\providecommand \natexlab [1]{#1}%
\providecommand \enquote  [1]{``#1''}%
\providecommand \bibnamefont  [1]{#1}%
\providecommand \bibfnamefont [1]{#1}%
\providecommand \citenamefont [1]{#1}%
\providecommand \href@noop [0]{\@secondoftwo}%
\providecommand \href [0]{\begingroup \@sanitize@url \@href}%
\providecommand \@href[1]{\@@startlink{#1}\@@href}%
\providecommand \@@href[1]{\endgroup#1\@@endlink}%
\providecommand \@sanitize@url [0]{\catcode `\\12\catcode `\$12\catcode
  `\&12\catcode `\#12\catcode `\^12\catcode `\_12\catcode `\%12\relax}%
\providecommand \@@startlink[1]{}%
\providecommand \@@endlink[0]{}%
\providecommand \url  [0]{\begingroup\@sanitize@url \@url }%
\providecommand \@url [1]{\endgroup\@href {#1}{\urlprefix }}%
\providecommand \urlprefix  [0]{URL }%
\providecommand \Eprint [0]{\href }%
\providecommand \doibase [0]{http://dx.doi.org/}%
\providecommand \selectlanguage [0]{\@gobble}%
\providecommand \bibinfo  [0]{\@secondoftwo}%
\providecommand \bibfield  [0]{\@secondoftwo}%
\providecommand \translation [1]{[#1]}%
\providecommand \BibitemOpen [0]{}%
\providecommand \bibitemStop [0]{}%
\providecommand \bibitemNoStop [0]{.\EOS\space}%
\providecommand \EOS [0]{\spacefactor3000\relax}%
\providecommand \BibitemShut  [1]{\csname bibitem#1\endcsname}%
\let\auto@bib@innerbib\@empty
\bibitem [{\citenamefont {Wen}(2004)}]{WenQFT2004}%
  \BibitemOpen
  \bibfield  {author} {\bibinfo {author} {\bibfnamefont {X.-G.}\ \bibnamefont
  {Wen}},\ }\href@noop {} {\emph {\bibinfo {title} {{Quantum Field Theory of
  Many-Body Systems}}}}\ (\bibinfo  {publisher} {Oxford University Press},\
  \bibinfo {address} {New York},\ \bibinfo {year} {2004})\BibitemShut {NoStop}%
\bibitem [{\citenamefont {Wan}\ \emph {et~al.}(2011)\citenamefont {Wan},
  \citenamefont {Turner}, \citenamefont {Vishwanath},\ and\ \citenamefont
  {Savrasov}}]{Ari2010}%
  \BibitemOpen
  \bibfield  {author} {\bibinfo {author} {\bibfnamefont {X.}~\bibnamefont
  {Wan}}, \bibinfo {author} {\bibfnamefont {A.}~\bibnamefont {Turner}},
  \bibinfo {author} {\bibfnamefont {A.}~\bibnamefont {Vishwanath}}, \ and\
  \bibinfo {author} {\bibfnamefont {S.~Y.}\ \bibnamefont {Savrasov}},\
  }\href@noop {} {\bibfield  {journal} {\bibinfo  {journal} {Physical Review
  B}\ }\textbf {\bibinfo {volume} {83}},\ \bibinfo {pages} {205101} (\bibinfo
  {year} {2011})}\BibitemShut {NoStop}%
\bibitem [{\citenamefont {Burkov}\ and\ \citenamefont
  {Balents}(2011)}]{Burkov2011a}%
  \BibitemOpen
  \bibfield  {author} {\bibinfo {author} {\bibfnamefont {A.~A.}\ \bibnamefont
  {Burkov}}\ and\ \bibinfo {author} {\bibfnamefont {L.}~\bibnamefont
  {Balents}},\ }\href@noop {} {\bibfield  {journal} {\bibinfo  {journal}
  {Physical Review Letters}\ }\textbf {\bibinfo {volume} {107}},\ \bibinfo
  {pages} {127205} (\bibinfo {year} {2011})}\BibitemShut {NoStop}%
\bibitem [{\citenamefont {Armitage}\ \emph {et~al.}(2018)\citenamefont
  {Armitage}, \citenamefont {Mele},\ and\ \citenamefont
  {Vishwanath}}]{Armitage2018}%
  \BibitemOpen
  \bibfield  {author} {\bibinfo {author} {\bibfnamefont {N.}~\bibnamefont
  {Armitage}}, \bibinfo {author} {\bibfnamefont {E.~J.}\ \bibnamefont {Mele}},
  \ and\ \bibinfo {author} {\bibfnamefont {A.}~\bibnamefont {Vishwanath}},\
  }\href@noop {} {\bibfield  {journal} {\bibinfo  {journal} {Reviews of Modern
  Physics}\ }\textbf {\bibinfo {volume} {90}},\ \bibinfo {pages} {15001}
  (\bibinfo {year} {2018})}\BibitemShut {NoStop}%
\bibitem [{\citenamefont {Fu}\ \emph {et~al.}(2007)\citenamefont {Fu},
  \citenamefont {Kane},\ and\ \citenamefont {Mele}}]{Fu2007}%
  \BibitemOpen
  \bibfield  {author} {\bibinfo {author} {\bibfnamefont {L.}~\bibnamefont
  {Fu}}, \bibinfo {author} {\bibfnamefont {C.~L.}\ \bibnamefont {Kane}}, \ and\
  \bibinfo {author} {\bibfnamefont {E.~J.}\ \bibnamefont {Mele}},\ }\href@noop
  {} {\bibfield  {journal} {\bibinfo  {journal} {Physical Review Letters}\
  }\textbf {\bibinfo {volume} {98}},\ \bibinfo {pages} {106803} (\bibinfo
  {year} {2007})}\BibitemShut {NoStop}%
\bibitem [{\citenamefont {Moore}\ and\ \citenamefont
  {Balents}(2007)}]{Moore2007}%
  \BibitemOpen
  \bibfield  {author} {\bibinfo {author} {\bibfnamefont {J.~E.}\ \bibnamefont
  {Moore}}\ and\ \bibinfo {author} {\bibfnamefont {L.}~\bibnamefont
  {Balents}},\ }\href {\doibase 10.1103/PhysRevB.75.121306} {\bibfield
  {journal} {\bibinfo  {journal} {Phys. Rev. B}\ }\textbf {\bibinfo {volume}
  {75}},\ \bibinfo {pages} {121306} (\bibinfo {year} {2007})}\BibitemShut
  {NoStop}%
\bibitem [{\citenamefont {Teo}\ \emph {et~al.}(2008)\citenamefont {Teo},
  \citenamefont {Fu},\ and\ \citenamefont {Kane}}]{Teo2008}%
  \BibitemOpen
  \bibfield  {author} {\bibinfo {author} {\bibfnamefont {J.~C.~Y.}\
  \bibnamefont {Teo}}, \bibinfo {author} {\bibfnamefont {L.}~\bibnamefont
  {Fu}}, \ and\ \bibinfo {author} {\bibfnamefont {C.~L.}\ \bibnamefont
  {Kane}},\ }\href@noop {} {\bibfield  {journal} {\bibinfo  {journal} {Physical
  Review B}\ }\textbf {\bibinfo {volume} {78}},\ \bibinfo {pages} {045426}
  (\bibinfo {year} {2008})}\BibitemShut {NoStop}%
\bibitem [{\citenamefont {Kane}\ and\ \citenamefont {Mele}(2005)}]{Kane2005}%
  \BibitemOpen
  \bibfield  {author} {\bibinfo {author} {\bibfnamefont {C.~L.}\ \bibnamefont
  {Kane}}\ and\ \bibinfo {author} {\bibfnamefont {E.~J.}\ \bibnamefont
  {Mele}},\ }\href@noop {} {\bibfield  {journal} {\bibinfo  {journal} {Physical
  Review Letters}\ }\textbf {\bibinfo {volume} {95}},\ \bibinfo {pages}
  {226801} (\bibinfo {year} {2005})}\BibitemShut {NoStop}%
\bibitem [{\citenamefont {Nikoli\'c}\ and\ \citenamefont
  {Tesanovi\'c}(2013{\natexlab{a}})}]{Nikolic2012b}%
  \BibitemOpen
  \bibfield  {author} {\bibinfo {author} {\bibfnamefont {P.}~\bibnamefont
  {Nikoli\'c}}\ and\ \bibinfo {author} {\bibfnamefont {Z.}~\bibnamefont
  {Tesanovi\'c}},\ }\href@noop {} {\bibfield  {journal} {\bibinfo  {journal}
  {Physical Review B}\ }\textbf {\bibinfo {volume} {87}},\ \bibinfo {pages}
  {134511} (\bibinfo {year} {2013}{\natexlab{a}})}\BibitemShut {NoStop}%
\bibitem [{\citenamefont {Nikoli\'c}\ \emph {et~al.}(2013)\citenamefont
  {Nikoli\'c}, \citenamefont {Duri\'c},\ and\ \citenamefont
  {Tesanovi\'c}}]{Nikolic2011a}%
  \BibitemOpen
  \bibfield  {author} {\bibinfo {author} {\bibfnamefont {P.}~\bibnamefont
  {Nikoli\'c}}, \bibinfo {author} {\bibfnamefont {T.}~\bibnamefont {Duri\'c}}, \
  and\ \bibinfo {author} {\bibfnamefont {Z.}~\bibnamefont {Tesanovi\'c}},\
  }\href@noop {} {\bibfield  {journal} {\bibinfo  {journal} {Physical Review
  Letters}\ }\textbf {\bibinfo {volume} {110}},\ \bibinfo {pages} {176804}
  (\bibinfo {year} {2013})}\BibitemShut {NoStop}%
\bibitem [{\citenamefont {Nikoli\'c}\ and\ \citenamefont
  {Tesanovi\'c}(2013{\natexlab{b}})}]{Nikolic2012a}%
  \BibitemOpen
  \bibfield  {author} {\bibinfo {author} {\bibfnamefont {P.}~\bibnamefont
  {Nikoli\'c}}\ and\ \bibinfo {author} {\bibfnamefont {Z.}~\bibnamefont
  {Tesanovi\'c}},\ }\href@noop {} {\bibfield  {journal} {\bibinfo  {journal}
  {Physical Review B}\ }\textbf {\bibinfo {volume} {87}},\ \bibinfo {pages}
  {104514} (\bibinfo {year} {2013}{\natexlab{b}})}\BibitemShut {NoStop}%
\bibitem [{\citenamefont {Nikoli\'c}(2016)}]{Nikolic2014a}%
  \BibitemOpen
  \bibfield  {author} {\bibinfo {author} {\bibfnamefont {P.}~\bibnamefont
  {Nikoli\'c}},\ }\href@noop {} {\bibfield  {journal} {\bibinfo  {journal}
  {Physical Review B}\ }\textbf {\bibinfo {volume} {94}},\ \bibinfo {pages}
  {064523} (\bibinfo {year} {2016})}\BibitemShut {NoStop}%
\bibitem [{\citenamefont {Fröhlich}\ and\ \citenamefont
  {Studer}(1992)}]{Frohlich1992}%
  \BibitemOpen
  \bibfield  {author} {\bibinfo {author} {\bibfnamefont {J.}~\bibnamefont
  {Fröhlich}}\ and\ \bibinfo {author} {\bibfnamefont {U.~M.}\ \bibnamefont
  {Studer}},\ }\href@noop {} {\bibfield  {journal} {\bibinfo  {journal}
  {Communications in Mathematical Physics}\ }\textbf {\bibinfo {volume}
  {148}},\ \bibinfo {pages} {553–600} (\bibinfo {year} {1992})}\BibitemShut
  {NoStop}%
\bibitem [{\citenamefont {Kang}\ \emph {et~al.}(2015)\citenamefont {Kang},
  \citenamefont {Kim}, \citenamefont {Kim}, \citenamefont {Kang}, \citenamefont
  {Denlinger},\ and\ \citenamefont {Min}}]{Kang2013}%
  \BibitemOpen
  \bibfield  {author} {\bibinfo {author} {\bibfnamefont {C.-J.}\ \bibnamefont
  {Kang}}, \bibinfo {author} {\bibfnamefont {J.}~\bibnamefont {Kim}}, \bibinfo
  {author} {\bibfnamefont {K.}~\bibnamefont {Kim}}, \bibinfo {author}
  {\bibfnamefont {J.-S.}\ \bibnamefont {Kang}}, \bibinfo {author}
  {\bibfnamefont {J.~D.}\ \bibnamefont {Denlinger}}, \ and\ \bibinfo {author}
  {\bibfnamefont {B.~I.}\ \bibnamefont {Min}},\ }\href@noop {} {\bibfield
  {journal} {\bibinfo  {journal} {Journal of the Physical Society of Japan}\
  }\textbf {\bibinfo {volume} {84}},\ \bibinfo {pages} {024722} (\bibinfo
  {year} {2015})}\BibitemShut {NoStop}%
\bibitem [{\citenamefont {Lu}\ \emph {et~al.}(2013)\citenamefont {Lu},
  \citenamefont {Zhao}, \citenamefont {Weng}, \citenamefont {Fang},\ and\
  \citenamefont {Dai}}]{Lu2013b}%
  \BibitemOpen
  \bibfield  {author} {\bibinfo {author} {\bibfnamefont {F.}~\bibnamefont
  {Lu}}, \bibinfo {author} {\bibfnamefont {J.}~\bibnamefont {Zhao}}, \bibinfo
  {author} {\bibfnamefont {H.}~\bibnamefont {Weng}}, \bibinfo {author}
  {\bibfnamefont {Z.}~\bibnamefont {Fang}}, \ and\ \bibinfo {author}
  {\bibfnamefont {X.}~\bibnamefont {Dai}},\ }\href@noop {} {\bibfield
  {journal} {\bibinfo  {journal} {Physical Review Letters}\ }\textbf {\bibinfo
  {volume} {110}},\ \bibinfo {pages} {096401} (\bibinfo {year}
  {2013})}\BibitemShut {NoStop}%
\bibitem [{\citenamefont {Nikoli\'c}(2014{\natexlab{a}})}]{Nikolic2014b}%
  \BibitemOpen
  \bibfield  {author} {\bibinfo {author} {\bibfnamefont {P.}~\bibnamefont
  {Nikoli\'c}},\ }\href@noop {} {\bibfield  {journal} {\bibinfo  {journal}
  {Physical Review B}\ }\textbf {\bibinfo {volume} {90}},\ \bibinfo {pages}
  {235107} (\bibinfo {year} {2014}{\natexlab{a}})}\BibitemShut {NoStop}%
\bibitem [{\citenamefont {Nakajima}\ \emph {et~al.}(2016)\citenamefont
  {Nakajima}, \citenamefont {Syers}, \citenamefont {Wang}, \citenamefont
  {Wang},\ and\ \citenamefont {Paglione}}]{Nakajima2016}%
  \BibitemOpen
  \bibfield  {author} {\bibinfo {author} {\bibfnamefont {Y.}~\bibnamefont
  {Nakajima}}, \bibinfo {author} {\bibfnamefont {P.}~\bibnamefont {Syers}},
  \bibinfo {author} {\bibfnamefont {X.}~\bibnamefont {Wang}}, \bibinfo {author}
  {\bibfnamefont {R.}~\bibnamefont {Wang}}, \ and\ \bibinfo {author}
  {\bibfnamefont {J.}~\bibnamefont {Paglione}},\ }\href@noop {} {\bibfield
  {journal} {\bibinfo  {journal} {Nature Physics}\ }\textbf {\bibinfo {volume}
  {12}},\ \bibinfo {pages} {213} (\bibinfo {year} {2016})},\ \bibinfo {note}
  {magnetoresistance hysteresis}\BibitemShut {NoStop}%
\bibitem [{\citenamefont {Lee}\ \emph {et~al.}(2016)\citenamefont {Lee},
  \citenamefont {Zhang}, \citenamefont {Liang}, \citenamefont {Fackler},
  \citenamefont {Yong}, \citenamefont {Wang}, \citenamefont {Paglione},
  \citenamefont {Greene},\ and\ \citenamefont {Takeuchi}}]{Takeuchi2016}%
  \BibitemOpen
  \bibfield  {author} {\bibinfo {author} {\bibfnamefont {S.}~\bibnamefont
  {Lee}}, \bibinfo {author} {\bibfnamefont {X.}~\bibnamefont {Zhang}}, \bibinfo
  {author} {\bibfnamefont {Y.}~\bibnamefont {Liang}}, \bibinfo {author}
  {\bibfnamefont {S.}~\bibnamefont {Fackler}}, \bibinfo {author} {\bibfnamefont
  {J.}~\bibnamefont {Yong}}, \bibinfo {author} {\bibfnamefont {X.}~\bibnamefont
  {Wang}}, \bibinfo {author} {\bibfnamefont {J.}~\bibnamefont {Paglione}},
  \bibinfo {author} {\bibfnamefont {R.~L.}\ \bibnamefont {Greene}}, \ and\
  \bibinfo {author} {\bibfnamefont {I.}~\bibnamefont {Takeuchi}},\ }\href@noop
  {} {\bibfield  {journal} {\bibinfo  {journal} {Physical Review X}\ }\textbf
  {\bibinfo {volume} {6}},\ \bibinfo {pages} {031031} (\bibinfo {year}
  {2016})}\BibitemShut {NoStop}%
\bibitem [{\citenamefont {Pesin}\ and\ \citenamefont
  {Balents}(2010)}]{Pesin2010}%
  \BibitemOpen
  \bibfield  {author} {\bibinfo {author} {\bibfnamefont {D.}~\bibnamefont
  {Pesin}}\ and\ \bibinfo {author} {\bibfnamefont {L.}~\bibnamefont
  {Balents}},\ }\href@noop {} {\bibfield  {journal} {\bibinfo  {journal}
  {Nature Physics}\ }\textbf {\bibinfo {volume} {6}},\ \bibinfo {pages} {376}
  (\bibinfo {year} {2010})}\BibitemShut {NoStop}%
\bibitem [{\citenamefont {Chen}\ and\ \citenamefont
  {Hermele}(2012)}]{Chen2012}%
  \BibitemOpen
  \bibfield  {author} {\bibinfo {author} {\bibfnamefont {G.}~\bibnamefont
  {Chen}}\ and\ \bibinfo {author} {\bibfnamefont {M.}~\bibnamefont {Hermele}},\
  }\href@noop {} {\bibfield  {journal} {\bibinfo  {journal} {Physical Review
  B}\ }\textbf {\bibinfo {volume} {86}},\ \bibinfo {pages} {235129} (\bibinfo
  {year} {2012})}\BibitemShut {NoStop}%
\bibitem [{\citenamefont {Ishii}\ \emph {et~al.}(2015)\citenamefont {Ishii},
  \citenamefont {Mizuta}, \citenamefont {Kato}, \citenamefont {Ozaki},
  \citenamefont {Weng},\ and\ \citenamefont {Onoda}}]{Ishii2015}%
  \BibitemOpen
  \bibfield  {author} {\bibinfo {author} {\bibfnamefont {F.}~\bibnamefont
  {Ishii}}, \bibinfo {author} {\bibfnamefont {Y.~P.}\ \bibnamefont {Mizuta}},
  \bibinfo {author} {\bibfnamefont {T.}~\bibnamefont {Kato}}, \bibinfo {author}
  {\bibfnamefont {T.}~\bibnamefont {Ozaki}}, \bibinfo {author} {\bibfnamefont
  {H.}~\bibnamefont {Weng}}, \ and\ \bibinfo {author} {\bibfnamefont
  {S.}~\bibnamefont {Onoda}},\ }\href {\doibase 10.7566/JPSJ.84.073703}
  {\bibfield  {journal} {\bibinfo  {journal} {Journal of the Physical Society
  of Japan}\ }\textbf {\bibinfo {volume} {84}},\ \bibinfo {pages} {073703}
  (\bibinfo {year} {2015})}\BibitemShut {NoStop}%
\bibitem [{\citenamefont {Kondo}\ \emph {et~al.}(2015)\citenamefont {Kondo},
  \citenamefont {Nakayama}, \citenamefont {Chen}, \citenamefont {Ishikawa},
  \citenamefont {Moon}, \citenamefont {Yamamoto}, \citenamefont {Ota},
  \citenamefont {Malaeb}, \citenamefont {Kanai}, \citenamefont {Nakashima},
  \citenamefont {Ishida}, \citenamefont {Yoshida}, \citenamefont {Yamamoto},
  \citenamefont {Matsunami}, \citenamefont {Kimura}, \citenamefont {Inami},
  \citenamefont {Ono}, \citenamefont {Kumigashira}, \citenamefont {Nakatsuji},
  \citenamefont {Balents},\ and\ \citenamefont {Shin}}]{Kondo2015}%
  \BibitemOpen
  \bibfield  {author} {\bibinfo {author} {\bibfnamefont {T.}~\bibnamefont
  {Kondo}}, \bibinfo {author} {\bibfnamefont {M.}~\bibnamefont {Nakayama}},
  \bibinfo {author} {\bibfnamefont {R.}~\bibnamefont {Chen}}, \bibinfo {author}
  {\bibfnamefont {J.}~\bibnamefont {Ishikawa}}, \bibinfo {author}
  {\bibfnamefont {E.-G.}\ \bibnamefont {Moon}}, \bibinfo {author}
  {\bibfnamefont {T.}~\bibnamefont {Yamamoto}}, \bibinfo {author}
  {\bibfnamefont {Y.}~\bibnamefont {Ota}}, \bibinfo {author} {\bibfnamefont
  {W.}~\bibnamefont {Malaeb}}, \bibinfo {author} {\bibfnamefont
  {H.}~\bibnamefont {Kanai}}, \bibinfo {author} {\bibfnamefont
  {Y.}~\bibnamefont {Nakashima}}, \bibinfo {author} {\bibfnamefont
  {Y.}~\bibnamefont {Ishida}}, \bibinfo {author} {\bibfnamefont
  {R.}~\bibnamefont {Yoshida}}, \bibinfo {author} {\bibfnamefont
  {H.}~\bibnamefont {Yamamoto}}, \bibinfo {author} {\bibfnamefont
  {M.}~\bibnamefont {Matsunami}}, \bibinfo {author} {\bibfnamefont
  {S.}~\bibnamefont {Kimura}}, \bibinfo {author} {\bibfnamefont
  {N.}~\bibnamefont {Inami}}, \bibinfo {author} {\bibfnamefont
  {K.}~\bibnamefont {Ono}}, \bibinfo {author} {\bibfnamefont {H.}~\bibnamefont
  {Kumigashira}}, \bibinfo {author} {\bibfnamefont {S.}~\bibnamefont
  {Nakatsuji}}, \bibinfo {author} {\bibfnamefont {L.}~\bibnamefont {Balents}},
  \ and\ \bibinfo {author} {\bibfnamefont {S.}~\bibnamefont {Shin}},\
  }\href@noop {} {\bibfield  {journal} {\bibinfo  {journal} {Nature
  Communications}\ }\textbf {\bibinfo {volume} {6}},\ \bibinfo {pages} {10042}
  (\bibinfo {year} {2015})}\BibitemShut {NoStop}%
\bibitem [{\citenamefont {Nakayama}\ \emph {et~al.}(2016)\citenamefont
  {Nakayama}, \citenamefont {Kondo}, \citenamefont {Tian}, \citenamefont
  {Ishikawa}, \citenamefont {Halim}, \citenamefont {Bareille}, \citenamefont
  {Malaeb}, \citenamefont {Kuroda}, \citenamefont {Tomita}, \citenamefont
  {Ideta}, \citenamefont {Tanaka}, \citenamefont {Matsunami}, \citenamefont
  {Kimura}, \citenamefont {Inami}, \citenamefont {Ono}, \citenamefont
  {Kumigashira}, \citenamefont {Balents}, \citenamefont {Nakatsuji},\ and\
  \citenamefont {Shin}}]{Nakayama2016}%
  \BibitemOpen
  \bibfield  {author} {\bibinfo {author} {\bibfnamefont {M.}~\bibnamefont
  {Nakayama}}, \bibinfo {author} {\bibfnamefont {T.}~\bibnamefont {Kondo}},
  \bibinfo {author} {\bibfnamefont {Z.}~\bibnamefont {Tian}}, \bibinfo {author}
  {\bibfnamefont {J.~J.}\ \bibnamefont {Ishikawa}}, \bibinfo {author}
  {\bibfnamefont {M.}~\bibnamefont {Halim}}, \bibinfo {author} {\bibfnamefont
  {C.}~\bibnamefont {Bareille}}, \bibinfo {author} {\bibfnamefont
  {W.}~\bibnamefont {Malaeb}}, \bibinfo {author} {\bibfnamefont
  {K.}~\bibnamefont {Kuroda}}, \bibinfo {author} {\bibfnamefont
  {T.}~\bibnamefont {Tomita}}, \bibinfo {author} {\bibfnamefont
  {S.}~\bibnamefont {Ideta}}, \bibinfo {author} {\bibfnamefont
  {K.}~\bibnamefont {Tanaka}}, \bibinfo {author} {\bibfnamefont
  {M.}~\bibnamefont {Matsunami}}, \bibinfo {author} {\bibfnamefont
  {S.}~\bibnamefont {Kimura}}, \bibinfo {author} {\bibfnamefont
  {N.}~\bibnamefont {Inami}}, \bibinfo {author} {\bibfnamefont
  {K.}~\bibnamefont {Ono}}, \bibinfo {author} {\bibfnamefont {H.}~\bibnamefont
  {Kumigashira}}, \bibinfo {author} {\bibfnamefont {L.}~\bibnamefont
  {Balents}}, \bibinfo {author} {\bibfnamefont {S.}~\bibnamefont {Nakatsuji}},
  \ and\ \bibinfo {author} {\bibfnamefont {S.}~\bibnamefont {Shin}},\ }\href
  {\doibase 10.1103/PhysRevLett.117.056403} {\bibfield  {journal} {\bibinfo
  {journal} {Physical Review Letters}\ }\textbf {\bibinfo {volume} {117}},\
  \bibinfo {pages} {056403} (\bibinfo {year} {2016})}\BibitemShut {NoStop}%
\bibitem [{\citenamefont {Zhang}\ \emph {et~al.}(2017)\citenamefont {Zhang},
  \citenamefont {Haule},\ and\ \citenamefont {Vanderbilt}}]{Zhang2017}%
  \BibitemOpen
  \bibfield  {author} {\bibinfo {author} {\bibfnamefont {H.}~\bibnamefont
  {Zhang}}, \bibinfo {author} {\bibfnamefont {K.}~\bibnamefont {Haule}}, \ and\
  \bibinfo {author} {\bibfnamefont {D.}~\bibnamefont {Vanderbilt}},\
  }\href@noop {} {\bibfield  {journal} {\bibinfo  {journal} {Physical Review
  Letters}\ }\textbf {\bibinfo {volume} {118}},\ \bibinfo {pages} {026404}
  (\bibinfo {year} {2017})}\BibitemShut {NoStop}%
\bibitem [{\citenamefont {Wang}\ \emph {et~al.}(2017)\citenamefont {Wang},
  \citenamefont {Go},\ and\ \citenamefont {Millis}}]{Wang2017}%
  \BibitemOpen
  \bibfield  {author} {\bibinfo {author} {\bibfnamefont {R.}~\bibnamefont
  {Wang}}, \bibinfo {author} {\bibfnamefont {A.}~\bibnamefont {Go}}, \ and\
  \bibinfo {author} {\bibfnamefont {A.~J.}\ \bibnamefont {Millis}},\
  }\href@noop {} {\bibfield  {journal} {\bibinfo  {journal} {Physical Review
  B}\ }\textbf {\bibinfo {volume} {95}},\ \bibinfo {pages} {045133} (\bibinfo
  {year} {2017})}\BibitemShut {NoStop}%
\bibitem [{\citenamefont {Cheng}\ \emph {et~al.}(2017)\citenamefont {Cheng},
  \citenamefont {Ohtsuki}, \citenamefont {Chaudhuri}, \citenamefont
  {Nakatsuji}, \citenamefont {Lippmaa},\ and\ \citenamefont
  {Armitage}}]{Cheng2017}%
  \BibitemOpen
  \bibfield  {author} {\bibinfo {author} {\bibfnamefont {B.}~\bibnamefont
  {Cheng}}, \bibinfo {author} {\bibfnamefont {T.}~\bibnamefont {Ohtsuki}},
  \bibinfo {author} {\bibfnamefont {D.}~\bibnamefont {Chaudhuri}}, \bibinfo
  {author} {\bibfnamefont {S.}~\bibnamefont {Nakatsuji}}, \bibinfo {author}
  {\bibfnamefont {M.}~\bibnamefont {Lippmaa}}, \ and\ \bibinfo {author}
  {\bibfnamefont {N.}~\bibnamefont {Armitage}},\ }\href {\doibase
  10.1038/s41467-017-02121-y} {\bibfield  {journal} {\bibinfo  {journal}
  {Nature Communications}\ }\textbf {\bibinfo {volume} {8}},\ \bibinfo {pages}
  {1} (\bibinfo {year} {2017})}\BibitemShut {NoStop}%
\bibitem [{\citenamefont {Shinaoka}\ \emph {et~al.}(2019)\citenamefont
  {Shinaoka}, \citenamefont {Motome}, \citenamefont {Miyake}, \citenamefont
  {Ishibashi},\ and\ \citenamefont {Werner}}]{Shinaoka2019}%
  \BibitemOpen
  \bibfield  {author} {\bibinfo {author} {\bibfnamefont {H.}~\bibnamefont
  {Shinaoka}}, \bibinfo {author} {\bibfnamefont {Y.}~\bibnamefont {Motome}},
  \bibinfo {author} {\bibfnamefont {T.}~\bibnamefont {Miyake}}, \bibinfo
  {author} {\bibfnamefont {S.}~\bibnamefont {Ishibashi}}, \ and\ \bibinfo
  {author} {\bibfnamefont {P.}~\bibnamefont {Werner}},\ }\href@noop {}
  {\bibfield  {journal} {\bibinfo  {journal} {Journal of Physics: Condensed
  Matter}\ }\textbf {\bibinfo {volume} {31}},\ \bibinfo {pages} {323001}
  (\bibinfo {year} {2019})}\BibitemShut {NoStop}%
\bibitem [{\citenamefont {Abrikosov}\ and\ \citenamefont
  {Beneslavskii}(1971)}]{Abrikosov1971}%
  \BibitemOpen
  \bibfield  {author} {\bibinfo {author} {\bibfnamefont {A.~A.}\ \bibnamefont
  {Abrikosov}}\ and\ \bibinfo {author} {\bibfnamefont {D.}~\bibnamefont
  {Beneslavskii}},\ }\href@noop {} {\bibfield  {journal} {\bibinfo  {journal}
  {Zhurnal Eksperimental'noi i Teoreticheskoi Fiziki}\ }\textbf {\bibinfo
  {volume} {59}},\ \bibinfo {pages} {1280–1298} (\bibinfo {year}
  {1971})}\BibitemShut {NoStop}%
\bibitem [{\citenamefont {Moon}\ \emph {et~al.}(2013)\citenamefont {Moon},
  \citenamefont {Xu}, \citenamefont {Kim},\ and\ \citenamefont
  {Balents}}]{Moon2013}%
  \BibitemOpen
  \bibfield  {author} {\bibinfo {author} {\bibfnamefont {E.-G.}\ \bibnamefont
  {Moon}}, \bibinfo {author} {\bibfnamefont {C.}~\bibnamefont {Xu}}, \bibinfo
  {author} {\bibfnamefont {Y.~B.}\ \bibnamefont {Kim}}, \ and\ \bibinfo
  {author} {\bibfnamefont {L.}~\bibnamefont {Balents}},\ }\href@noop {}
  {\bibfield  {journal} {\bibinfo  {journal} {Physical Review Letters}\
  }\textbf {\bibinfo {volume} {111}},\ \bibinfo {pages} {206401} (\bibinfo
  {year} {2013})}\BibitemShut {NoStop}%
\bibitem [{\citenamefont {Nandkishore}\ and\ \citenamefont
  {Parameswaran}(2017)}]{Parameswaran2017}%
  \BibitemOpen
  \bibfield  {author} {\bibinfo {author} {\bibfnamefont {R.~M.}\ \bibnamefont
  {Nandkishore}}\ and\ \bibinfo {author} {\bibfnamefont {S.~A.}\ \bibnamefont
  {Parameswaran}},\ }\href {\doibase 10.1103/PhysRevB.95.205106} {\bibfield
  {journal} {\bibinfo  {journal} {Physical Review B}\ }\textbf {\bibinfo
  {volume} {95}},\ \bibinfo {pages} {205106} (\bibinfo {year}
  {2017})}\BibitemShut {NoStop}%
\bibitem [{\citenamefont {Herbut}\ and\ \citenamefont
  {Janssen}(2014)}]{Herbut2014}%
  \BibitemOpen
  \bibfield  {author} {\bibinfo {author} {\bibfnamefont {I.~F.}\ \bibnamefont
  {Herbut}}\ and\ \bibinfo {author} {\bibfnamefont {L.}~\bibnamefont
  {Janssen}},\ }\href {\doibase 10.1103/PhysRevLett.113.106401} {\bibfield
  {journal} {\bibinfo  {journal} {Physical Review Letters}\ }\textbf {\bibinfo
  {volume} {113}},\ \bibinfo {pages} {106401} (\bibinfo {year}
  {2014})}\BibitemShut {NoStop}%
\bibitem [{\citenamefont {Janssen}\ and\ \citenamefont
  {Herbut}(2015)}]{Herbut2015}%
  \BibitemOpen
  \bibfield  {author} {\bibinfo {author} {\bibfnamefont {L.}~\bibnamefont
  {Janssen}}\ and\ \bibinfo {author} {\bibfnamefont {I.~F.}\ \bibnamefont
  {Herbut}},\ }\href {\doibase 10.1103/PhysRevB.92.045117} {\bibfield
  {journal} {\bibinfo  {journal} {Physical Review B}\ }\textbf {\bibinfo
  {volume} {92}},\ \bibinfo {pages} {045117} (\bibinfo {year}
  {2015})}\BibitemShut {NoStop}%
\bibitem [{\citenamefont {Janssen}\ and\ \citenamefont
  {Herbut}(2016)}]{Herbut2016b}%
  \BibitemOpen
  \bibfield  {author} {\bibinfo {author} {\bibfnamefont {L.}~\bibnamefont
  {Janssen}}\ and\ \bibinfo {author} {\bibfnamefont {I.~F.}\ \bibnamefont
  {Herbut}},\ }\href {\doibase 10.1103/PhysRevB.93.165109} {\bibfield
  {journal} {\bibinfo  {journal} {Physical Review B}\ }\textbf {\bibinfo
  {volume} {93}},\ \bibinfo {pages} {165109} (\bibinfo {year}
  {2016})}\BibitemShut {NoStop}%
\bibitem [{\citenamefont {Janssen}\ and\ \citenamefont
  {Herbut}(2017)}]{Herbut2017}%
  \BibitemOpen
  \bibfield  {author} {\bibinfo {author} {\bibfnamefont {L.}~\bibnamefont
  {Janssen}}\ and\ \bibinfo {author} {\bibfnamefont {I.~F.}\ \bibnamefont
  {Herbut}},\ }\href {\doibase 10.1103/PhysRevB.95.075101} {\bibfield
  {journal} {\bibinfo  {journal} {Physical Review B}\ }\textbf {\bibinfo
  {volume} {95}},\ \bibinfo {pages} {075101} (\bibinfo {year}
  {2017})}\BibitemShut {NoStop}%
\bibitem [{\citenamefont {Boettcher}\ and\ \citenamefont
  {Herbut}(2017)}]{Herbut2017b}%
  \BibitemOpen
  \bibfield  {author} {\bibinfo {author} {\bibfnamefont {I.}~\bibnamefont
  {Boettcher}}\ and\ \bibinfo {author} {\bibfnamefont {I.~F.}\ \bibnamefont
  {Herbut}},\ }\href {\doibase 10.1103/PhysRevB.95.075149} {\bibfield
  {journal} {\bibinfo  {journal} {Physical Review B}\ }\textbf {\bibinfo
  {volume} {95}},\ \bibinfo {pages} {075149} (\bibinfo {year}
  {2017})}\BibitemShut {NoStop}%
\bibitem [{\citenamefont {Boettcher}\ and\ \citenamefont
  {Herbut}(2016)}]{Herbut2016}%
  \BibitemOpen
  \bibfield  {author} {\bibinfo {author} {\bibfnamefont {I.}~\bibnamefont
  {Boettcher}}\ and\ \bibinfo {author} {\bibfnamefont {I.~F.}\ \bibnamefont
  {Herbut}},\ }\href {\doibase 10.1103/PhysRevB.93.205138} {\bibfield
  {journal} {\bibinfo  {journal} {Physical Review B}\ }\textbf {\bibinfo
  {volume} {93}},\ \bibinfo {pages} {205138} (\bibinfo {year}
  {2016})}\BibitemShut {NoStop}%
\bibitem [{\citenamefont {Roy}\ \emph {et~al.}(2019)\citenamefont {Roy},
  \citenamefont {Ghorashi}, \citenamefont {Foster},\ and\ \citenamefont
  {Nevidomskyy}}]{Nevidomskyy2019}%
  \BibitemOpen
  \bibfield  {author} {\bibinfo {author} {\bibfnamefont {B.}~\bibnamefont
  {Roy}}, \bibinfo {author} {\bibfnamefont {S.~A.~A.}\ \bibnamefont
  {Ghorashi}}, \bibinfo {author} {\bibfnamefont {M.~S.}\ \bibnamefont
  {Foster}}, \ and\ \bibinfo {author} {\bibfnamefont {A.~H.}\ \bibnamefont
  {Nevidomskyy}},\ }\href {\doibase 10.1103/PhysRevB.99.054505} {\bibfield
  {journal} {\bibinfo  {journal} {Physical Review B}\ }\textbf {\bibinfo
  {volume} {99}},\ \bibinfo {pages} {054505} (\bibinfo {year}
  {2019})}\BibitemShut {NoStop}%
\bibitem [{\citenamefont {Sato}\ and\ \citenamefont {Ando}(2017)}]{Sato2017}%
  \BibitemOpen
  \bibfield  {author} {\bibinfo {author} {\bibfnamefont {M.}~\bibnamefont
  {Sato}}\ and\ \bibinfo {author} {\bibfnamefont {Y.}~\bibnamefont {Ando}},\
  }\href {\doibase 10.1088/1361-6633/aa6ac7} {\bibfield  {journal} {\bibinfo
  {journal} {Reports on Progress in Physics}\ }\textbf {\bibinfo {volume}
  {80}},\ \bibinfo {pages} {076501} (\bibinfo {year} {2017})}\BibitemShut
  {NoStop}%
\bibitem [{\citenamefont {Schnyder}\ \emph {et~al.}(2008)\citenamefont
  {Schnyder}, \citenamefont {Ryu}, \citenamefont {Furusaki},\ and\
  \citenamefont {Ludwig}}]{Schnyder2008}%
  \BibitemOpen
  \bibfield  {author} {\bibinfo {author} {\bibfnamefont {A.~P.}\ \bibnamefont
  {Schnyder}}, \bibinfo {author} {\bibfnamefont {S.}~\bibnamefont {Ryu}},
  \bibinfo {author} {\bibfnamefont {A.}~\bibnamefont {Furusaki}}, \ and\
  \bibinfo {author} {\bibfnamefont {A.~W.~W.}\ \bibnamefont {Ludwig}},\
  }\href@noop {} {\bibfield  {journal} {\bibinfo  {journal} {Physical Review
  B}\ }\textbf {\bibinfo {volume} {78}},\ \bibinfo {pages} {195125} (\bibinfo
  {year} {2008})}\BibitemShut {NoStop}%
\bibitem [{\citenamefont {Matsuura}\ \emph {et~al.}(2013)\citenamefont
  {Matsuura}, \citenamefont {Chang}, \citenamefont {Schnyder},\ and\
  \citenamefont {Ryu}}]{Schnyder2013}%
  \BibitemOpen
  \bibfield  {author} {\bibinfo {author} {\bibfnamefont {S.}~\bibnamefont
  {Matsuura}}, \bibinfo {author} {\bibfnamefont {P.-Y.}\ \bibnamefont {Chang}},
  \bibinfo {author} {\bibfnamefont {A.~P.}\ \bibnamefont {Schnyder}}, \ and\
  \bibinfo {author} {\bibfnamefont {S.}~\bibnamefont {Ryu}},\ }\href@noop {}
  {\bibfield  {journal} {\bibinfo  {journal} {New Journal of Physics}\ }\textbf
  {\bibinfo {volume} {15}},\ \bibinfo {pages} {065001} (\bibinfo {year}
  {2013})}\BibitemShut {NoStop}%
\bibitem [{\citenamefont {Chiu}\ and\ \citenamefont
  {Schnyder}(2014)}]{Schnyder2014}%
  \BibitemOpen
  \bibfield  {author} {\bibinfo {author} {\bibfnamefont {C.-K.}\ \bibnamefont
  {Chiu}}\ and\ \bibinfo {author} {\bibfnamefont {A.~P.}\ \bibnamefont
  {Schnyder}},\ }\href {\doibase 10.1103/PhysRevB.90.205136} {\bibfield
  {journal} {\bibinfo  {journal} {Physical Review B}\ }\textbf {\bibinfo
  {volume} {90}},\ \bibinfo {pages} {205136} (\bibinfo {year}
  {2014})}\BibitemShut {NoStop}%
\bibitem [{\citenamefont {Shiozaki}\ and\ \citenamefont
  {Sato}(2014)}]{Shiozaki2014}%
  \BibitemOpen
  \bibfield  {author} {\bibinfo {author} {\bibfnamefont {K.}~\bibnamefont
  {Shiozaki}}\ and\ \bibinfo {author} {\bibfnamefont {M.}~\bibnamefont
  {Sato}},\ }\href {\doibase 10.1103/PhysRevB.90.165114} {\bibfield  {journal}
  {\bibinfo  {journal} {Physical Review B}\ }\textbf {\bibinfo {volume} {90}},\
  \bibinfo {pages} {165114} (\bibinfo {year} {2014})}\BibitemShut {NoStop}%
\bibitem [{\citenamefont {Norman}(1995)}]{Norman1995}%
  \BibitemOpen
  \bibfield  {author} {\bibinfo {author} {\bibfnamefont {M.~R.}\ \bibnamefont
  {Norman}},\ }\href {\doibase 10.1103/PhysRevB.52.15093} {\bibfield  {journal}
  {\bibinfo  {journal} {Physical Review B}\ }\textbf {\bibinfo {volume} {52}},\
  \bibinfo {pages} {15093–15094} (\bibinfo {year} {1995})}\BibitemShut
  {NoStop}%
\bibitem [{\citenamefont {Sato}(2006)}]{Sato2006}%
  \BibitemOpen
  \bibfield  {author} {\bibinfo {author} {\bibfnamefont {M.}~\bibnamefont
  {Sato}},\ }\href {\doibase 10.1103/PhysRevB.73.214502} {\bibfield  {journal}
  {\bibinfo  {journal} {Physical Review B}\ }\textbf {\bibinfo {volume} {73}},\
  \bibinfo {pages} {214502} (\bibinfo {year} {2006})}\BibitemShut {NoStop}%
\bibitem [{\citenamefont {Roy}(2008)}]{Roy2008}%
  \BibitemOpen
  \bibfield  {author} {\bibinfo {author} {\bibfnamefont {R.}~\bibnamefont
  {Roy}},\ }\href@noop {} {\  (\bibinfo {year} {2008})},\ \bibinfo {note}
  {arXiv:0803.2868}\BibitemShut {NoStop}%
\bibitem [{\citenamefont {Qi}\ \emph {et~al.}(2009)\citenamefont {Qi},
  \citenamefont {Hughes}, \citenamefont {Raghu},\ and\ \citenamefont
  {Zhang}}]{Qi2009b}%
  \BibitemOpen
  \bibfield  {author} {\bibinfo {author} {\bibfnamefont {X.-L.}\ \bibnamefont
  {Qi}}, \bibinfo {author} {\bibfnamefont {T.~L.}\ \bibnamefont {Hughes}},
  \bibinfo {author} {\bibfnamefont {S.}~\bibnamefont {Raghu}}, \ and\ \bibinfo
  {author} {\bibfnamefont {S.-C.}\ \bibnamefont {Zhang}},\ }\href@noop {}
  {\bibfield  {journal} {\bibinfo  {journal} {Physical Review Letters}\
  }\textbf {\bibinfo {volume} {102}},\ \bibinfo {pages} {187001} (\bibinfo
  {year} {2009})}\BibitemShut {NoStop}%
\bibitem [{\citenamefont {Micklitz}\ and\ \citenamefont
  {Norman}(2009)}]{Norman2009}%
  \BibitemOpen
  \bibfield  {author} {\bibinfo {author} {\bibfnamefont {T.}~\bibnamefont
  {Micklitz}}\ and\ \bibinfo {author} {\bibfnamefont {M.~R.}\ \bibnamefont
  {Norman}},\ }\href {\doibase 10.1103/PhysRevB.80.100506} {\bibfield
  {journal} {\bibinfo  {journal} {Physical Review B}\ }\textbf {\bibinfo
  {volume} {80}},\ \bibinfo {pages} {100506} (\bibinfo {year}
  {2009})}\BibitemShut {NoStop}%
\bibitem [{\citenamefont {Sato}\ and\ \citenamefont
  {Fujimoto}(2009)}]{Fujimoto2009}%
  \BibitemOpen
  \bibfield  {author} {\bibinfo {author} {\bibfnamefont {M.}~\bibnamefont
  {Sato}}\ and\ \bibinfo {author} {\bibfnamefont {S.}~\bibnamefont
  {Fujimoto}},\ }\href {\doibase 10.1103/PhysRevB.79.094504} {\bibfield
  {journal} {\bibinfo  {journal} {Physical Review B}\ }\textbf {\bibinfo
  {volume} {79}},\ \bibinfo {pages} {094504} (\bibinfo {year}
  {2009})}\BibitemShut {NoStop}%
\bibitem [{\citenamefont {Tanaka}\ \emph {et~al.}(2009)\citenamefont {Tanaka},
  \citenamefont {Yokoyama}, \citenamefont {Balatsky},\ and\ \citenamefont
  {Nagaosa}}]{Tanaka2009b}%
  \BibitemOpen
  \bibfield  {author} {\bibinfo {author} {\bibfnamefont {Y.}~\bibnamefont
  {Tanaka}}, \bibinfo {author} {\bibfnamefont {T.}~\bibnamefont {Yokoyama}},
  \bibinfo {author} {\bibfnamefont {A.~V.}\ \bibnamefont {Balatsky}}, \ and\
  \bibinfo {author} {\bibfnamefont {N.}~\bibnamefont {Nagaosa}},\ }\href
  {\doibase 10.1103/PhysRevB.79.060505} {\bibfield  {journal} {\bibinfo
  {journal} {Physical Review B}\ }\textbf {\bibinfo {volume} {79}},\ \bibinfo
  {pages} {060505} (\bibinfo {year} {2009})}\BibitemShut {NoStop}%
\bibitem [{\citenamefont {Kobayashi}\ \emph {et~al.}(2014)\citenamefont
  {Kobayashi}, \citenamefont {Shiozaki}, \citenamefont {Tanaka},\ and\
  \citenamefont {Sato}}]{Kobayashi2014}%
  \BibitemOpen
  \bibfield  {author} {\bibinfo {author} {\bibfnamefont {S.}~\bibnamefont
  {Kobayashi}}, \bibinfo {author} {\bibfnamefont {K.}~\bibnamefont {Shiozaki}},
  \bibinfo {author} {\bibfnamefont {Y.}~\bibnamefont {Tanaka}}, \ and\ \bibinfo
  {author} {\bibfnamefont {M.}~\bibnamefont {Sato}},\ }\href {\doibase
  10.1103/PhysRevB.90.024516} {\bibfield  {journal} {\bibinfo  {journal}
  {Physical Review B}\ }\textbf {\bibinfo {volume} {90}},\ \bibinfo {pages}
  {024516} (\bibinfo {year} {2014})}\BibitemShut {NoStop}%
\bibitem [{\citenamefont {Yang}\ \emph {et~al.}(2014)\citenamefont {Yang},
  \citenamefont {Pan},\ and\ \citenamefont {Zhang}}]{Zhang2014}%
  \BibitemOpen
  \bibfield  {author} {\bibinfo {author} {\bibfnamefont {S.~A.}\ \bibnamefont
  {Yang}}, \bibinfo {author} {\bibfnamefont {H.}~\bibnamefont {Pan}}, \ and\
  \bibinfo {author} {\bibfnamefont {F.}~\bibnamefont {Zhang}},\ }\href
  {\doibase 10.1103/PhysRevLett.113.046401} {\bibfield  {journal} {\bibinfo
  {journal} {Physical Review Letters}\ }\textbf {\bibinfo {volume} {113}},\
  \bibinfo {pages} {046401} (\bibinfo {year} {2014})}\BibitemShut {NoStop}%
\bibitem [{\citenamefont {Zhao}\ \emph {et~al.}(2016)\citenamefont {Zhao},
  \citenamefont {Schnyder},\ and\ \citenamefont {Wang}}]{Zhao2016}%
  \BibitemOpen
  \bibfield  {author} {\bibinfo {author} {\bibfnamefont {Y.~X.}\ \bibnamefont
  {Zhao}}, \bibinfo {author} {\bibfnamefont {A.~P.}\ \bibnamefont {Schnyder}},
  \ and\ \bibinfo {author} {\bibfnamefont {Z.~D.}\ \bibnamefont {Wang}},\
  }\href {\doibase 10.1103/PhysRevLett.116.156402} {\bibfield  {journal}
  {\bibinfo  {journal} {Physical Review Letters}\ }\textbf {\bibinfo {volume}
  {116}},\ \bibinfo {pages} {156402} (\bibinfo {year} {2016})}\BibitemShut
  {NoStop}%
\bibitem [{\citenamefont {Micklitz}\ and\ \citenamefont
  {Norman}(2017)}]{Norman2017}%
  \BibitemOpen
  \bibfield  {author} {\bibinfo {author} {\bibfnamefont {T.}~\bibnamefont
  {Micklitz}}\ and\ \bibinfo {author} {\bibfnamefont {M.~R.}\ \bibnamefont
  {Norman}},\ }\href {\doibase 10.1103/PhysRevLett.118.207001} {\bibfield
  {journal} {\bibinfo  {journal} {Physical Review Letters}\ }\textbf {\bibinfo
  {volume} {118}},\ \bibinfo {pages} {207001} (\bibinfo {year}
  {2017})}\BibitemShut {NoStop}%
\bibitem [{\citenamefont {Nomoto}\ and\ \citenamefont
  {Ikeda}(2017)}]{Ikeda2017}%
  \BibitemOpen
  \bibfield  {author} {\bibinfo {author} {\bibfnamefont {T.}~\bibnamefont
  {Nomoto}}\ and\ \bibinfo {author} {\bibfnamefont {H.}~\bibnamefont {Ikeda}},\
  }\href {\doibase 10.7566/jpsj.86.023703} {\bibfield  {journal} {\bibinfo
  {journal} {Journal of the Physical Society of Japan}\ }\textbf {\bibinfo
  {volume} {86}},\ \bibinfo {pages} {023703} (\bibinfo {year}
  {2017})}\BibitemShut {NoStop}%
\bibitem [{\citenamefont {Béri}(2010)}]{Beri2010}%
  \BibitemOpen
  \bibfield  {author} {\bibinfo {author} {\bibfnamefont {B.}~\bibnamefont
  {Béri}},\ }\href {\doibase 10.1103/PhysRevB.81.134515} {\bibfield  {journal}
  {\bibinfo  {journal} {Physical Review B}\ }\textbf {\bibinfo {volume} {81}},\
  \bibinfo {pages} {134515} (\bibinfo {year} {2010})}\BibitemShut {NoStop}%
\bibitem [{\citenamefont {Nikoli\'c}(2014{\natexlab{b}})}]{Nikolic2014}%
  \BibitemOpen
  \bibfield  {author} {\bibinfo {author} {\bibfnamefont {P.}~\bibnamefont
  {Nikoli\'c}},\ }\href@noop {} {\bibfield  {journal} {\bibinfo  {journal}
  {Physical Review A}\ }\textbf {\bibinfo {volume} {90}},\ \bibinfo {pages}
  {023623} (\bibinfo {year} {2014}{\natexlab{b}})}\BibitemShut {NoStop}%
\bibitem [{\citenamefont {Lu}\ and\ \citenamefont {Wang}(2008)}]{Wang2008b}%
  \BibitemOpen
  \bibfield  {author} {\bibinfo {author} {\bibfnamefont {H.-Y.}\ \bibnamefont
  {Lu}}\ and\ \bibinfo {author} {\bibfnamefont {Q.-H.}\ \bibnamefont {Wang}},\
  }\href@noop {} {\bibfield  {journal} {\bibinfo  {journal} {Chinese Physics
  Letters}\ }\textbf {\bibinfo {volume} {25}},\ \bibinfo {pages} {3746}
  (\bibinfo {year} {2008})}\BibitemShut {NoStop}%
\bibitem [{\citenamefont {Kashuba}\ and\ \citenamefont
  {Fal’ko}(2009)}]{Kashuba2009}%
  \BibitemOpen
  \bibfield  {author} {\bibinfo {author} {\bibfnamefont {O.}~\bibnamefont
  {Kashuba}}\ and\ \bibinfo {author} {\bibfnamefont {V.~I.}\ \bibnamefont
  {Fal’ko}},\ }\href@noop {} {\bibfield  {journal} {\bibinfo  {journal}
  {Physical Review B}\ }\textbf {\bibinfo {volume} {80}},\ \bibinfo {pages}
  {241404 (R)} (\bibinfo {year} {2009})}\BibitemShut {NoStop}%
\bibitem [{\citenamefont {Nikoli\'c}\ \emph {et~al.}(2022)\citenamefont
  {Nikoli\'c}, \citenamefont {Xu}, \citenamefont {Ohtsuki}, \citenamefont
  {Nakatsuji},\ and\ \citenamefont {Drichko}}]{Nikolic2022a}%
  \BibitemOpen
  \bibfield  {author} {\bibinfo {author} {\bibfnamefont {P.}~\bibnamefont
  {Nikoli\'c}}, \bibinfo {author} {\bibfnamefont {Y.}~\bibnamefont {Xu}},
  \bibinfo {author} {\bibfnamefont {T.}~\bibnamefont {Ohtsuki}}, \bibinfo
  {author} {\bibfnamefont {S.}~\bibnamefont {Nakatsuji}}, \ and\ \bibinfo
  {author} {\bibfnamefont {N.}~\bibnamefont {Drichko}},\ }\href {\doibase
  10.48550/ARXIV.2204.13722} {\  (\bibinfo {year} {2022}),\
  10.48550/ARXIV.2204.13722},\ \bibinfo {note} {arXiv:2204.13722}\BibitemShut
  {NoStop}%
\end{thebibliography}


%


\end{document}